\documentclass[aps, 10pt, prr, twocolumn, amssymb, amsmath, amsfonts, showpacs, superscriptaddress, preprintnumbers, a4paper, longbibliography]{revtex4-2}
\usepackage{amsmath, amssymb, amsthm}
\usepackage{times}
\usepackage{xcolor}
\usepackage{graphicx}
\usepackage[english]{babel}
\usepackage[
  breaklinks,
  colorlinks=true,
  bookmarks=true,
  bookmarksopenlevel=3,
  bookmarksnumbered=true,
  menucolor=blue,
  linkcolor=blue,
  citecolor=blue,
  urlcolor=blue,
  pdfcreator={GhostScript},
  pdfkeywords={},
  pdfstartview=FitH]{hyperref}
\begin{document}

\def\lptms{Universit\'e Paris-Saclay, CNRS, LPTMS, 91405, Orsay, France.}
\def\lps{Universit\'e Paris-Saclay, CNRS, Laboratoire de Physique des Solides, 91405, Orsay, France.}

\title{Pairing in spinless fermions and spin chains with next-nearest neighbour interactions}

\author{Lorenzo Gotta}\email{lorenzo.gotta@u-psud.fr}\affiliation{\lptms}
\author{Leonardo Mazza}\affiliation{\lptms}
\author{Pascal Simon}\affiliation{\lps}
\author{Guillaume Roux}\affiliation{\lptms}

\date{\today}

\begin{abstract}

We investigate the phase diagrams of a one-dimensional lattice model of fermions and of a spin chain with interactions extending up to next-nearest neighbour range. 
In particular, we investigate the appearance of regions with dominant pairing physics in the presence of nearest-neighbour and next-nearest-neighbour interactions.
Our analysis is based on analytical calculations in the classical limit, bosonization techniques and large-scale density-matrix renormalization group numerical simulations.
The phase diagram, which is investigated in all relevant filling regimes, displays a remarkably rich collection of phases, including Luttinger liquids, phase separation, charge-density waves, bond-order phases, and exotic cluster Luttinger liquids with paired particles. 
In relation with recent studies, we show several emergent transition lines with a central charge $c=3/2$ between the Luttinger-liquid and the cluster Luttinger liquid phases. These results could be experimentally investigated using highly-tunable quantum simulators.

\end{abstract}

\maketitle

\section{Introduction}

Whether or not unpaired Majorana zero modes could appear in one-dimensional systems without the coupling to an external superconducting device, is a question that has recently triggered a lot of interest.
Since the mean-field approach represented by Kitaev's chain~\cite{kitaev_unpaired_2001} cannot answer the query because of the low dimensionality of the setup,
an increasing recent literature has focused on pairing phenomena in number-conserving {one-dimensional} systems~\cite{ortiz_many-body_2014, keselman_gapless_2015, lang_topological_2015, iemini_localized_2015, ruhman_topological_2015, kells_multiparticle_2015, ruhman_topological_2017, Kane2017,iemini_majorana_2017}.  
{As a main result of this research subject, it has been proposed that the} use of inhomogeneous systems where paired phases are coupled to normal phases is then expected to reveal zero-energy Majorana boundary modes, which should be pinned at the interfaces~\cite{ruhman_topological_2015}.
{For this reason, the study of pairing physics in one-dimensional fermionic setups has recently become an extremely interesting topic and this observation constitutes the broad motivation of the present work.}

A paradigmatic model for investigating pairing physics is a one-dimensional lattice model of fermions with density-density interactions which extend up to next-nearest neighbours (NNN)~\cite{poilblanc_insulator-metal_1997, zhuravlev_electronic_1997, zhuravlev_breakdown_2000, zhuravlev_one-dimensional_2001, ejima_phase_2005, duan_bond-order_2011, hohenadler_interaction-range_2012, li_ground_2019,Ren_2012,Patrick_2019}.
Some parts of its phase diagrams have already been studied and identified as paired phases, which appear both for repulsive~\cite{mattioli_cluster_2013, dalmonte_cluster_2015} and attractive~\cite{Kane2017,he_emergent_2019} interactions. 
A systematic analysis has been only presented at half filling, but limited to the fully-repulsive case~\cite{mishra_phase_2011}. 
Extensions to longer-range interactions have been the focus of Ref.~[\onlinecite{szyniszewski_fermionic_2018}].
In general, the employed techniques are various, and 
the study of one-dimensional setups naturally benefits from the possibility of using ad-hoc field-theory methods such as bosonization~\cite{capponi_effects_2000, cazalilla_bosonizing_2004, giamarchi_quantum_2010}, or numerical tensor-network approaches based on matrix-product states~\cite{schollwock_density-matrix_2005, schollwock_density-matrix_2011}.

In this article we present a comprehensive study of the general structure of the phase diagram of the aforementioned model using both approximate analytical treatments and extensive numerical simulations.
Our study reproduces the mentioned known results, and extends the analysis to  a larger parameter space focusing on four representative densities, $n=\frac 15$, $\frac 13$, $\frac 25$ and  $\frac 12$. 
Our discussion highlights that the phase diagram enriches progressively as the density is increased.
Obviously, we put a particular emphasis on pairing phenomena, which appear both for attractive and repulsive interactions.
% We show that when the NNN interaction is repulsive, pairing phenomena take place both for repulsive and for attractive nearest-neighbour interactions.
By completely mapping out the phase diagram, we expect to significantly ease the future search for Majorana fermions in number-conserving systems.

We want to highlight two significant points of our work.
The former is that
the study of paired phases is particularly interesting also because the transition from a standard Luttinger liquid requires a non-trivial bosonisation description~\cite{Kane2017}. 
Indeed, together with standard Luttinger liquid, one speaks of pair cluster Luttinger liquids (CLL), with gapped single-particle excitations and gapless pair degrees of freedom.
Our numerical simulations access the transition without using perturbative arguments and reproduce some expected results, such as the low-energy effective theory at the transition, which is a conformal field theory with central charge $c=1+\frac 12$. This point is discussed in the main text.
  
% Luttinger-liquid theory is a cornerstone of our understanding of many-body quantum physics in one dimension.
% According to it, they display important collective properties that differ significantly from that of the standard Fermi liquid theory that is routinely employed for interacting fermionic gases in higher dimensions.
% Luttinger-liquid theory provides a solid starting point for the study of perturbed systems, and even of higher-dimensional setups with the so-called coupled-wire approach.

The latter is that the scope of the article goes beyond the search for Majorana fermions in electronic systems. 
Thanks to the Jordan-Wigner mapping, our results can be easily recast in spin language and provide insights into the physics of arrays of Rydberg atoms.
Recent experiments have shown that
it is now possible to organize
individual atoms according to periodic arrays of microscopic dipole traps separated by few micrometers~\cite{nogrette_single-atom_2014, labuhn_single-atom_2014, lee_three-dimensional_2016, barredo_atom-by-atom_2016, endres_atom-by-atom_2016, nguyen_towards_2018, cooper_alkaline-earth_2018, madjarov_atomic-array_2019, cortinas_laser_2020}.
The excitation of such trapped atoms to a Rydberg state~\cite{gallagher_rydberg_1994, sibalic_rydberg_2018} characterized by a strong electronic dipole ensures that atoms interact notwithstanding their distances, and this
has produced a setup which is an almost paradigmatic realisation of a quantum simulator for quantum spin models~\cite{weimer_rydberg_2010, labuhn_tunable_2016, de_leseleuc_accurate_2018, browaeys_many-body_2020}.
In some special regimes, the realised model is an instance of our Hamiltonian.

The paper is organized  as follows. In Sec.~\ref{first} we present the explicit form of the model Hamiltonian both in the fermionic and in the spin formulation, we indicate the parameters chosen for the numerical density-matrix renormalization group (DMRG) investigations and the physical observables employed throughout the article.
Sec.~\ref{second} is fully devoted to the analysis of the phase diagram at density $n<\frac{1}{3}$, where the most interesting finding is a pair CLL in the attractive part of the phase diagram. 
The work continues in Sec.~\ref{third} with the characterization of a charge-density wave (CDW) in the strong coupling repulsive regime of the model at density $n=\frac{1}{3}$. 
The content of Sec.~\ref{fourth} is devoted to the density regime $\frac{1}{3}<n<\frac{1}{2}$. We focus both on the robustness of the pair CLL phase observed at lower density and on the observation of a novel CLL phase in the repulsive regime. 
As a last step, we provide in Sec.~\ref{fifth} the main features of the phase diagram structure at density $n=\frac{1}{2}$, where commensurability effects are responsible for the disappearance of the liquid behaviour observed at lower densities in favor of insulating CDW phases.

% We show that it is separated from the standard weak-coupling Luttinger liquid (LL) phase by a conformal critical point with central charge $c=\frac{3}{2}$ and is thoroughly characterized by means of suitable density and correlation decay properties.

% We characterize the resulting crystalline order, that can also be predicted on the basis of simple classical considerations.
% , by monitoring spectral and entropic properties and correlation function signatures, thereby demonstrating their compatibility with a CDW phase.

% We investigate both phases by monitoring the same observables employed in Sec.~\ref{first}, so that we can differentiate them on the basis of qualitative features that are reinterpreted in light of classical limit properties.

% We further demonstrate how the crystal phase induced by the NNN interaction is preceded at intermediate coupling by the so called bond order (BO) phase, featuring localization of the charge on the bond between sites, and which signatures accompany the transition from the LL phase to phase separation (PS).

\section{Model, methods and overview of the phase diagrams} \label{first}
\subsection{Hamiltonian}

The Hamiltonian that we are going to characterize is defined on a lattice of $L$ sites and admits two equivalent formulations in terms of fermionic or spin degrees of freedom. 
The fermionic model reads:
\begin{equation}
H=\sum_{j=1}^L\left[-t(\hat{c}^{\dag}_{j}\hat{c}_{j+1}+h.c.)+U_{1}\hat{n}_{j}\hat{n}_{j+1}+U_{2}\hat{n}_{j}\hat{n}_{j+2}\right], \label{fermionic}
\end{equation} 
where $\hat{c}_{j},\hat{c}^{\dag}_{j}$ are fermionic creation and annihilation operators satisfying the canonical anticommutation relations  $\{\hat{c}_{i},\hat{c}_{j}\}=0$ and $\{\hat{c}_{i},\hat{c}^{\dag}_{j}\}=\delta_{i,j}$, $\hat{n}_{j}=\hat{c}^{\dag}_{j}\hat{c}_{j}$ is the number operator at site $j$, $t$ denotes the hopping amplitude (we set $t=1$ in the rest of the paper) and $U_{1},U_{2}$ represent, respectively, the strength of the nearest neighbour (NN) and NNN density-density interactions. In the fermionic formulation \eqref{fermionic}, which we are going to refer to in the rest of the work, the model Hamiltonian that we consider describes fermions on a $1D$ lattice interacting via a soft-shoulder potential with interaction range $r_{c}=2$.
In what follows, we study the zero-temperature properties of  \eqref{fermionic} for real $U_{1}$ and positive $U_{2}\geq 0$ at a given density $n=\frac{1}{L}\sum_{j}\langle\hat{n}_{j}\rangle = N/L$ with $N$ the fixed total number of particles.

In order to switch from the fermionic to the spin Hamiltonian,  we use the Jordan-Wigner transformation
\begin{equation}
\begin{cases}
\hat{c}_{j}=\prod\limits_{l=1}^{j-1}e^{-i\pi\left(\hat{S}_{l}^{z}+\frac{1}{2}\right)}\hat{S}_{j}^{-},\\
%\hat{c}^{\dag}_{j}=\hat{S}_{j}^{+}\prod_{l=1}^{j-1}e^{i\pi\left(\hat{S}_{l}^{z}+\frac{1}{2}\right)},\\
\hat{c}^{\dag}_{j}\hat{c}_{j}=\hat{S}_{j}^{z}+\frac{1}{2},
\end{cases}
\end{equation}
where $\hat{S}_{j}^{k}$ with $k\in\{x,y,z\}$ are spin $1/2$ operators, 
leading to the following spin Hamiltonian:
\begin{equation}
H=\sum_{j}\left[-2t\left(\hat{S}^{+}_{j}\hat{S}^{-}_{j+1}+h.c.\right)+U_{1}\hat{S}^{z}_{j}\hat{S}^{z}_{j+1}+U_{2}\hat{S}^{z}_{j}\hat{S}^{z}_{j+2}\right], \label{spin}
\end{equation}
where we have dropped constant terms and terms proportional to the full magnetization, which commutes with the Hamiltonian. It thus corresponds to the well-known XXZ spin chain model with an extra antiferromagnetic NNN Ising term.
Such Hamiltonian could be realized in quantum simulators gathering a collection of interacting two-level systems, such as Rydberg simulators~\cite{browaeys_many-body_2020}.

\subsection{Particle-hole symmetry}

Thanks to particle-hole transformation $\hat{c}_{j}\rightarrow \hat{c}^{\dag}_{j}$, we can restrict our study to densities in the interval $0\leq n \leq 1/2$. 
Indeed, it translates to $n\rightarrow 1-n$, under which the Hamiltonian transforms as:
\begin{equation}
H(t,U_{1},U_{2})\rightarrow H(-t,U_{1},U_{2})+(U_{1}+U_{2})(L-2N)\,.
\end{equation}
By further noticing that one can remove the minus sign in front of the hopping parameter by means of the unitary transformation $\hat{c}_j\rightarrow -\hat{c}_j$ on even sites $j$, one can readily prove that the behaviour of the holes in the $n>\frac{1}{2}$ regime coincides with that of the particles at density $n'=1-n$ with the same interaction parameters.
% {\color{red}We thus restrict our study to densities $n \leq \frac{1}{2}$.}

\subsection{Numerical details and numerically-computed quantities}

Complementary to analytical tools, we carry out numerical simulations using the DMRG algorithm, a state-of-the-art method to tackle 1D systems with short range interactions~\cite{White1992,White1993,schollwock_density-matrix_2005}. We use two implementations: a traditional one, and one based on matrix product states using the ITensor library~\cite{ITensor}.
We use both periodic boundary conditions (PBC) and open boundary conditions (OBC), working with lattice sizes up to $L=80$ (resp. $L=140$), while keeping up to $m=2800$ states per block. For observables that depend on the particle statistics, we specify that all simulations are carried out on the fermionic model \eqref{fermionic}. Last, notice that systems with OBC may not exactly match the density definition $n=N/L$ because adding extra particles that fit to the edges is preferable to stabilize the expected density in the bulk and prevents the system from forming a defect in the bulk. Such a choice depends on the region of the phase diagram and on the nature of the underlying leading order.

\subsubsection{The phase diagrams}

We compute several quantities in order to map out the phase diagrams. 
First, the gapped nature of the low-energy excitations is inferred from the single-particle gap $\Delta_{1}$, while pairing occurs when the two-particle gap $\Delta_{2}$ vanishes. Numerically, finite-size gaps derived from ground-state energies $E_0(N,L)$ differences following
\begin{equation} \label{gap}
\Delta_{p}(L)=E_0(N+p,L)+E_0(N-p,L)-2E_0(N,L)\,
\end{equation}
are extrapolated with system size $L$ for $p=1,2$.

Another indication of the critical behaviour of the system emerges when computing the ground state energy density curvature $\partial_{s}^{2}\epsilon_{GS}$.
The latter quantity is defined as the directional second derivative of the ground state energy density $\epsilon_{GS}=\frac{E_0(N,L)}{L}$ along a curve $\gamma : \mathbb{R}\rightarrow\mathbb{R}^{2},\,\gamma(s)=\left(U_{1}(s),U_{2}(s)\right)$ in the $U_{1}-U_{2}$ parameter space:
\begin{equation} 
\label{curvature}
\partial_{s}^{2}\epsilon_{GS}=\frac{d^{2}}{ds^{2}}\epsilon_{GS}\left(\gamma(s)\right)\,.
\end{equation}
Indeed, its non-analyticities should signal zero-temperature quantum phase transitions.

A third probe to monitor critical phases and their central charge is the bipartite von Neumann entanglement entropy~\cite{holzhey_geometric_1994,calabrese_entanglement_2004}:
\begin{equation} \label{entanglement_entropy}
S_{A}=-\text{Tr}\left[\rho_{A}\log\rho_{A}\right],
\end{equation}
where $\rho_{A}$ is the reduced density matrix of subsystem $A$ with respect to the whole system. The central charge $c$ is estimated by fitting the finite size profile using the Cardy-Calabrese formula:
\begin{equation} 
\label{cardy-calabrese}
S_{L}(l)=\frac{c}{\alpha}\log\left[\frac{\beta L}{\pi}\sin\left(\pi\frac{l}{L}\right)\right]+C,
\end{equation}
where $l$ is the size of the left block $A$ length, $C$ represents a nonuniversal constant, and we have $\alpha=3, \beta=1$ for PBC. Additional oscillations in $S_{L}(l)$ are taken into account from the local kinetic energy profile~\cite{Laflorencie_2006, Affleck_2009, Roux_2009, Cardy_2010} by adding to eq. (\ref{cardy-calabrese}) a term of the form $B\langle c^{\dag}_{l}c_{l+1}+h.c. \rangle$, where $B$ is to be treated as a fitting parameter.

For each considered density $n$, we find particularly useful to present a sketch of the phase diagrams by plotting the bipartite entanglement entropy $S_L(L/2)$, whose peaks are an effective guide to the eye for phase transitions. Each phase diagram is analyzed with steps of $\Delta U_i = 0.125$; black lines are the phase transitions in the classical limit $t=0$ and red lines are lines for which we present additional numerical data.

\subsubsection{Observables}

In order to elucidate the nature of the CLL phases and their irreducibility to a standard LL phase, we introduce the Fourier transform $\delta n(k)$ of the density fluctuations $\langle\delta\hat{n}_{j}\rangle=\langle\hat{n}_{j}-n\rangle$, and the structure factor $S(k)$: 
% as the Fourier transform of the density-density correlation function $\langle\hat{n}_{j}\hat{n}_{j+r}\rangle$, i.e.:
\begin{align}
&\delta n(k)=\sum_{j=1}^{L} \langle\delta\hat{n}_{j}\rangle e^{-i\frac{(j-1)k}{L}}; \label{density_fourier} \\
&S(k)=\sum_{j=1}^{L}\left[\langle \hat{n}_{1}\hat{n}_{j}\rangle-\langle\hat{n}_{1}\rangle\langle\hat{n}_{j}\rangle\right] e^{-i\frac{(j-1)k}{L}}\,. \label{structure_factor}
\end{align} 
These observables are typically extracted from PBC simulations to minimize boundary effects. 

Additionally, we compute the decay of the single-particle correlation function and of the pair correlation functions 
\begin{equation}
G(r)=\langle\hat{c}^{\dag}_{j}\hat{c}_{j+r}\rangle, \qquad
P(r)=\langle\hat{c}^{\dag}_{j}\hat{c}^{\dag}_{j+1}\hat{c}_{j+r}\hat{c}_{j+r+1}\rangle ;
\end{equation}
the formulas reported here assume translational invariance of the problem (PBC).
In order to generically evaluate the enhancement of pairing fluctuations, we compute the so called average pair kinetic energy:
\begin{equation}
K_{P}=\frac{1}{L}\sum_{j}\langle\hat{c}^{\dag}_{j}\hat{c}^{\dag}_{j+1}\hat{c}_{j+2}\hat{c}_{j+3}+h.c.\rangle,
\end{equation}  
quantifying the magnitude of the NN pair hopping processes.

The phase diagram at $n=\frac 12$ requires some further theoretical tools.
The BO phase is identified via the BO parameter:
\begin{equation} \label{bond_order}
O^{\text{BO}}=\frac{1}{L}\sum_{j=1}^{L}(-1)^{j}\langle{c}^{\dag}_{j}\hat{c}_{j+1}+h.c.\rangle,
\end{equation}
and we characterize the $U_{2}$-induced CDW order by computing the following CDW order parameter:
\begin{equation} \label{CDW_order}
O^{\text{CDW}}_j =\langle \hat{n}_{j+2}\rangle-\langle\hat{n}_{j}\rangle,
\end{equation}
in the bulk of the system, so that the unavoidable boundary effects are controlled as much as possible. 

\subsection{Summary of results}

The aim of the present work consists in giving a comprehensive view of the zero-temperature ground state properties of the model in \eqref{fermionic}, thereby highlighting the wealth of exotic collective behaviours achievable by varying interactions and density.
Our main results are summarized in Fig. \ref{fig:PDn02},  Fig. \ref{fig:PDn03}, Fig. \ref{fig:PDn04} and  Fig. \ref{fig:PDn05} where we present the phase diagrams for $n=\frac{1}{5}$, $n=\frac{1}{3}$, $n=\frac{2}{5}$ and $n=\frac{1}{2}$ respectively.

We start in Sec.~\ref{second} with the analysis of the low density regime, $n= \frac 15$ , whose global features are expected to be simpler than at higher densities due to the absence of commensurability and frustration effects.
The standard LL phase occupies the whole NN repulsive part of the phase diagram, $U_1>0$ (see Fig. \ref{fig:PDn02}). The attractive part $U_1<0$, instead, comprises three different phases: LL, PS and CLL. LL appears for weak NN interaction, $U_1 \sim 0$, whereas
PS appears in the limit of dominant $U_{1}<0$ term.

The CLL phase is interpreted as a phase of pairs where pairing effects dominate.
The transition between the CLL and the LL phases is clearly signaled by its entanglement properties, and in particular by a central charge $c=3/2$. This supports the interpretation of the quantum phase transition in terms of an emerging Ising degree of freedom~\cite{mattioli_cluster_2013, dalmonte_cluster_2015,Kane2017,ruhman_topological_2017,he_emergent_2019}, i.e. the critical behavior of the system along the transition belongs to the 2D Ising universality class.
This exotic liquid is further probed through its low energy spectral properties: the single-particle gap is non-zero, whereas the pair gap vanishes. 
% This strong coupling pairing regime is not connected to the weakly interacting  LL phase, which is clearly demonstrated by the density profile Fourier transform or the density structure factor.
In the strong-coupling regime $t \ll U_1,U_2$, the CLL phase is continuously connected to the classical limit description of the corresponding region of the phase diagram (the impossibility of a direct connection with the weakly-interacting LL is demonstrated by observables such as $\delta n(k)$ and $S(k)$).

The discussion continues in Sec.~\ref{third} with the case $n=\frac 13$~\cite{szyniszewski_fermionic_2018} where the phase diagram is presented in Fig. \ref{fig:PDn03}.
The study of the classical limit $t=0$ proposes that the attractive regime of the phase diagram $U_1<0$ coincides with that characterized at $n= \frac{1}{5}$. The numerical analysis mainly confirms this expectation, apart from a highly correlated phase at $U_1 \sim 0$.
On the other hand, the classical analysis shows a modification in the repulsive side $U_1>0$ and explains the onset of a CDW order with one particle each three sites.
The emergence of this CDW is captured by the opening of the single-particle gap across the transition, as well as by the onset of exponentially-decaying single-particle and pair correlators. The analysis of density fluctuations $\delta n(k)$ corroborates this interpretation.

% shows a peak in the Fourier transform of the density profile that matches the quasimomentum expected in the classical limit.

Sec.~\ref{fourth} focuses on a typical density lying in the interval $1/3<n<1/2$, with the case $n= \frac 25$~\cite{mattioli_cluster_2013, dalmonte_cluster_2015}. 
We first confirm the persistence of the CLL phase established at density $n=\frac 15$ for $U_1<0$ using an analysis analogous to that of section \ref{first}; qualitatively, for $U_1<0$ the phase diagram (see Fig. \ref{fig:PDn04}) coincides with that obtained at $n= \frac 13$. 
The repulsive regime $U_1>0$
is particularly rich.
We demonstrate that the standard LL phase survives in the region $U_{2}<\frac{U_{1}}{2}$ by providing the scaling of observables such as the central charge and the single-particle gap.
% we thus rule out the presence of a critical behaviour at intermediate to strong interaction strengths.
For $U_2> \frac{U_1}{2}$, we identify a transition from the LL phase to frustration-induced CLL phase. After commenting on its analogy with its attractive regime counterpart, we relate the position of the peaks in the density profile Fourier transform and structure factor to the classical limit cluster structure of the ground state of the system, while it is incompatible with both the LL phase and the attractive CLL expectations.

Sec.~\ref{fifth} is devoted to the description of the  phase diagram at half filling (see Fig. \ref{fig:PDn05}) with $n= \frac 12$~\cite{mishra_phase_2011}. We discuss how the combination of commensurable effects and interactions favors gapped phases, in addition to the other phases found at lower densities.
In particular, we develop a weak coupling bosonization treatment of the model explaining the qualitative properties of the gapped phases. By means of suitable order parameters, we distinguish a CDW phase induced by $U_{1}>0$ from a BO phase induced by $U_{2}>0$.
We conclude the treatment of the half-filled case by discriminating the BO phase from the repulsive NNN-induced CDW phase using a finite size scaling analysis of the order parameters of each phases. This shows that the BO phase acts as an intermediate coupling precursor of the CDW phase. Close to phase separation, and similarly to lower fillings, we show an enhancement of the pairing fluctuations as well as a divergence of the Luttinger parameter reminiscent of the CLL regime.

We show four phase diagrams for each typical density through the manuscript. The goal is to provide an overview of our current understanding of the relevant phases occurring when one varies the two interaction terms. We do not have a full description of the thermodynamic limit for all transitions lines. For the most important transitions, we have carried out a detailed analysis along a few cuts (in dashed red on the diagrams). The strong-coupling limit also provides a few reference lines (in dashed black on the diagrams). Some transitions, such as phase separation and the LL to CLL transition do not show strong finite-size effects, while for others (of the Kosterlitz-Thouless type for instance), finite-size effects are stronger. For each diagram, we chose to plot as a background the bulk entropy on a finite size system, with yellow corresponding to high values and dark blue to small values, close to zero. Its magnitude has no particular meaning but we found that its variation across the diagram nicely pictured the different phases and transitions lines, in particular for the LL to CLL we are interested in. For other boundaries, we use results from the literature when it exists, or propose lines that are guide to the eyes (in blue on the diagrams). We believe such diagrams, although not complete, are helpful for any further studies on the model, having in mind that each already represents 10,000 runs of DMRG.

\section{Phase diagram for $n=\frac 15$} \label{second}

\begin{figure}[b]
\centering
\includegraphics[width=0.95\columnwidth,clip]{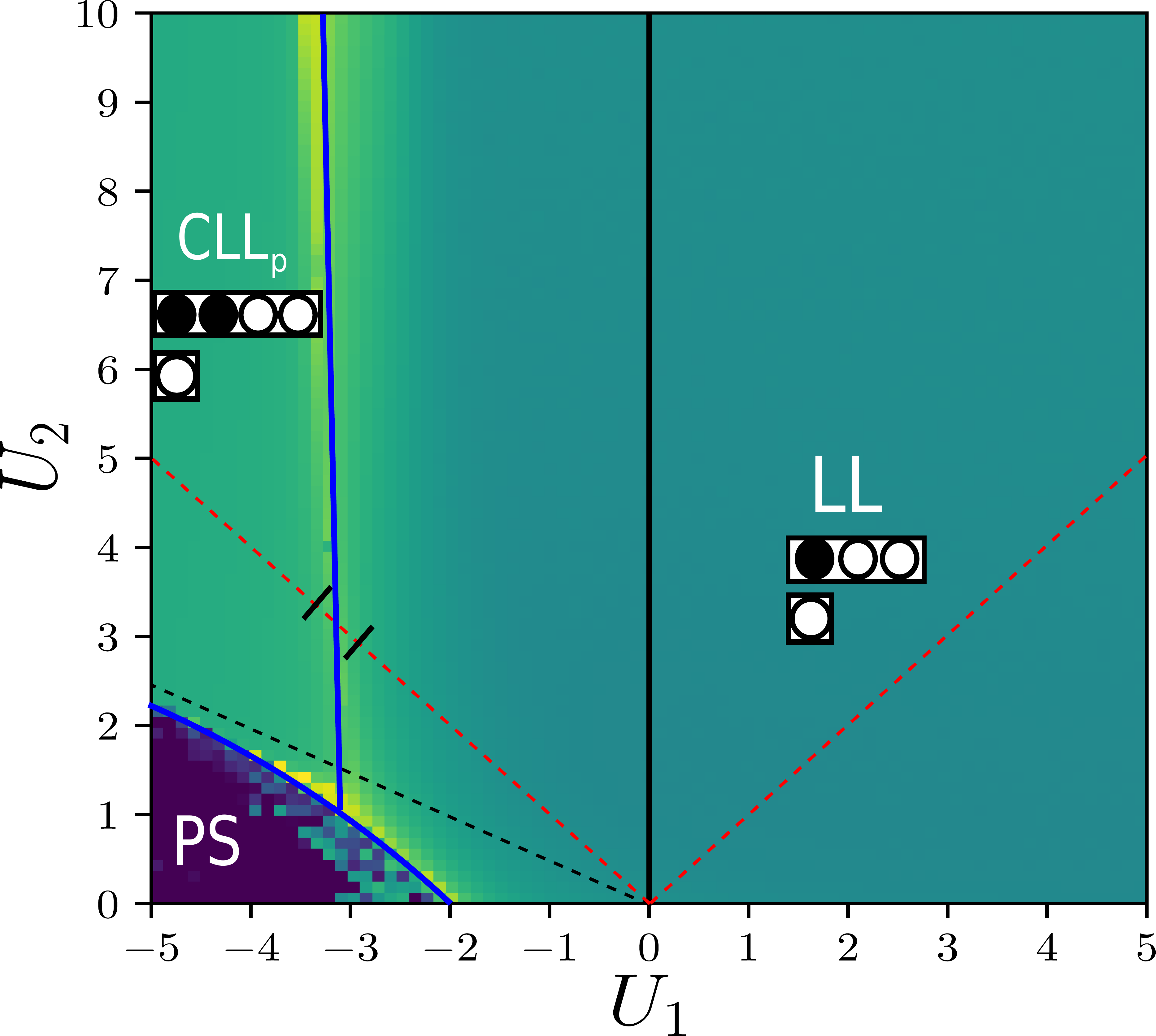}
\caption{{Phase diagram for $n=\frac{1}{5}$.} Color map for the background displays the entanglement entropy on $L=70$ chain with PBC. Black lines are classical transition lines obtained neglecting quantum fluctuation. Additional numerical simulations are presented for the points lying on the red lines. Blue lines are a guide to the eye for the main phase boundaries.}
\label{fig:PDn02}
\end{figure}

The main result discussed in this Section is the emergence of an exotic CLL phase whose fundamental gapless degrees of freedom are tightly bound pairs. This result is completely intuitive, given that it appears for strong and attractive $U_1$;
yet, its extension in the phase diagram and the nature of the transition to the weak-interacting LL phase are non-trivial and need an accurate characterization. 
The phase diagram obtained by  considerations based on the entanglement entropy is reported in Fig.~\ref{fig:PDn02}.

% In section~\ref{fourth}, we discuss its existence in the case of strongly repulsive interaction.

\subsection{Classical limit and Luttinger liquid approach}

We start with the description of the $U_2=0$, along which the system reduces to the XXZ model, see Eq.~\eqref{spin}. Phase separation (that is a ferromagnetic behaviour in spin language) occurs when $U_1 < -2$, while a LL phase covers the $U_1>0$ regime. In order to stabilize pairs, one intuitively requires the presence of a strong and attractive NN interaction, but have just observed that alone it yields to phase separation. Therefore, a repulsive $U_2$ is necessary in order to prevent the pairs from collapsing close to each other.
The low filling value is expected to avoid the emergence of commensurability effects and to enhance a dilute liquid of pairs.

Before turning our attention to the numerical data, we first obtain a qualitative understanding of the possible scenarios by studying the classical limit $t=0$ and applying strong-coupling arguments both for $U_{1}<0$ and $U_{1}>0$. 

When $U_{1}>0$ and $t=0$, the two competing ground state configurations are (i) the phase-separated one, which is a product state of $N$ occupied sites surrounded by empty sites, and (ii) the configurations with pairs separated by at least two empty sites. The latter is formally described by all possible permutations of blocks of type $A$ ($\bullet\bullet\circ\circ$) and blocks $B$ ($\circ$), where black circles refer to occupied sites and white circles represent empty sites. The numbers $N_{A}$ (resp. $N_{B}$) of block $A$ (resp. $B$) are solely determined by the total number of fermions and the size of the system since
\begin{align}
\begin{cases}
2N_{A}=N, \\
4N_{A}+N_{B}=L,
\end{cases}
\end{align}
so that $N_{A}=N/2$ and $N_{B}=L-2N$. The degeneracy of the paired configuration is given by $d=\frac{(N_{A}+N_{B})!}{N_{A}!N_{B}!}$, whose scaling is exponential in $L$.

The energy of a paired configuration equals $U_1$, and given that there are $N_A$ pairs, the ground-state energy density is $U_1 \frac{N_A}{L}=\frac{U_{1}}{2}n$. The phase-separated configuration, instead, has energy density $(U_{1}+U_{2})n$. As a result, the ground state energy density of the phase-separated configuration is optimal when $U_{2}<-\frac{U_{1}}{2}$.
This is the asymptotic classical limit separation line between the PS and CLL phases, and it is plotted in black in Fig.~\ref{fig:PDn02}.
When $U_{2}>-\frac{U_{1}}{2}$, we expect the highly-degenerate subspace of paired state to evolve into a liquid of tightly bound pairs once quantum fluctuations are reintroduced for $t\neq 0$. 

This result is further supported by a strong-coupling argument.
Assuming that, in this limit, the relevant dynamics takes place in the degenerate subspace of paired states described by $A$ and $B$ blocks only, we map the system onto an effective spin model of magnetization $M=N_{A}-N_{B}$ by associating each block $A$ (resp. $B$) with a spin-up (resp. spin-down).
Then, standard degenerate perturbation theory~\cite{mattioli_cluster_2013,dalmonte_cluster_2015,szyniszewski_generalized_2015} yields the following effective Hamiltonian
\begin{equation}
H=-\frac{J}{2}\sum_{j}\left[\hat{\mathcal{S}}^{+}_{j}\hat{\mathcal{S}}^{-}_{j+1}+h.c.\right]+J\Delta\sum_{j}\hat{\mathcal{S}}^{z}_{j}\hat{\mathcal{S}}^{z}_{j+1},
\end{equation}  
in which the $\mathcal{S}^\alpha$ are effective spin operators for the blocks and we drop the constant terms.
This XXZ model has effective couplings $J=\frac{2t^{2}}{U_{2}+|U_{1}|},\Delta=\frac{U_{2}}{2U_{2}+|U_{1}|}$ with an anisotropy parameter $\Delta\in (0,1)$.
In such regime, the effective XXZ chain is in the gapless LL regime, described by a $c=1$ conformal field theory. 
Consequently, the qualitative picture for the strong-coupling regime of the CLL phase is a Luttinger liquid of pairs that map hard-core bosons living on bonds.

Turning our attention to the purely repulsive interaction regime ($U_{1}>0,U_{2}>0$), the degenerate ground state subspace in the classical limit is the subspace generated by the basis states described as a sequence of blocks  $C$ ($\bullet\circ\circ$) and blocks $B$. The line $U_1=0$ thus constitutes another classical phase-transition line, and it is plotted in black in Fig.~\ref{fig:PDn02}.

It is easy to see that the system size and filling constraints impose $N_{B}=L-3N,N_{C}=N$, whereas, by performing a similar mapping to an effective spin model, the resulting effective Hamiltonian reads
\begin{equation} \label{XX}
H\simeq -t\sum_{j}\left[\hat{\mathcal{S}}_{j}^{\dag}\hat{\mathcal{S}}_{j+1}^{-}+h.c.\right]\,.
\end{equation}
Again, this XX model is described at low energies by a $c=1$ conformal field theory. 
As the fundamental granularity of the classical configurations comprises single particles, the strong-coupling limit is expected to be effectively adiabatically connected to the weak-coupling LL regime.
At low densities, such short range interactions, will never be able to drive the system to an instability towards nontrivial phases induced by frustration or commensurability effects. 
This claim that the LL phases extends over the whole repulsive region will be supported by numerical calculations.

Last, we recall the usual LL treatment of the weak-coupling regime stemming from the non-interacting point $U_1=U_2=0$. 
Bosonization maps the lattice operators to long-wavelength field operators $\psi_{R}(x),\psi_{L}(x)$ through $c_{j}\sim\sqrt{a}\left[\psi_{R}(ja)e^{ik_{F}ja}+\psi_{L}(ja)e^{-ik_{F}ja}\right]$ ($a$ being the lattice spacing and $k_F=\frac{\pi n}{a}$ being the Fermi wave-vector) and then re-expresses the latter as a function of two canonically conjugate bosonic fields $\phi (x),\partial_{x}\theta(x)$ satisfying $[\phi (x),\partial_{x'}\theta(x')]=i\delta (x-x')$.
The resulting effective Hamiltonian capturing the low energy properties of the system is the celebrated LL Hamiltonian~\cite{giamarchi_quantum_2010}:
\begin{equation} \label{LL_field_theory}
H=\frac{v}{2\pi}\int dx\left[\frac{1}{K}(\partial_{x}\phi)^{2}+K(\pi\partial_{x}\theta)^{2}\right],
\end{equation}
where $K$ denotes the Luttinger parameter and $v$ is the sound velocity of the gapless, linearly dispersing, collective density excitation modes. Such theory develops algebraic correlations parameterized by the $K$ parameter that, from pertubative calculations, reads: 
\begin{equation}
K(U_1,U_2;n) = \frac{1}{\sqrt{1+\frac{U_{1}\left[1-\cos(2\pi n)\right]+U_{2}\left[1-4\cos(4\pi n)\right]}{\pi \sin(\pi n)}}} \;.
\end{equation}

\subsection{Numerical results in the attractive regime}

\begin{figure}[t]
\centering
\includegraphics[width=\columnwidth]{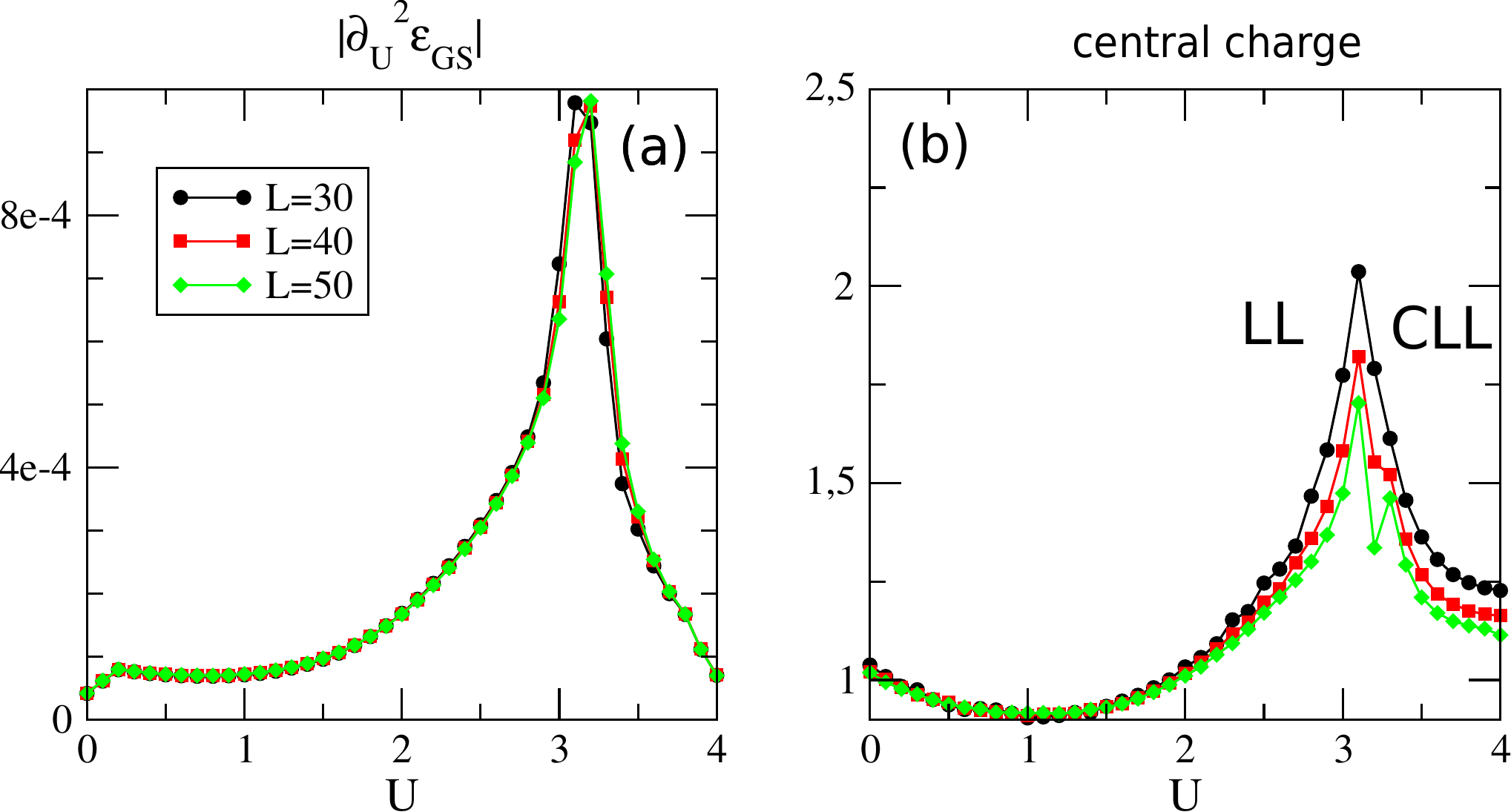}
\caption{DMRG results for $n=\frac{1}{5}$ along the $U=U_{2}=-U_{1}$ line of Fig.~\ref{fig:PDn02}.
 \textsf{\textbf{(a)}} ground state energy density curvature \eqref{curvature} for various sizes. The apparent non-analytic behavior suggests a critical point located in the range $3.1\leq U \leq 3.2$. 
\textsf{\textbf{(b)}} Extrapolated central charge from \eqref{cardy-calabrese}.
The numerics are compatible with a critical point with $c=3 / 2$ surrounded by the two $c=1$ LL and CLL phases.}
\label{fig:free_energy_c_neg_02}
\end{figure} 
%\subsubsection{Characterization of the transition}

In order to highlight the appearance of a transition to a different phase of matter, we focus on the line $U_2=-U_1$, which is plotted in red in Fig.~\ref{fig:PDn02}.
In Fig.~\ref{fig:free_energy_c_neg_02}(a), we present the profile of the second derivative of the ground-state energy density along this path. Despite the fact that it is an intensive quantity, it shows a non-analytical behavior that we interpret as an indication of the presence of the critical point, bearing in mind the fact that the numerical data alone do not allow to discriminate between a cusp and a genuine divergence.

The existence of a transition is further supported by the plot of the central charge along the very same line, as shown in Fig.~\ref{fig:free_energy_c_neg_02}(b), where a $c=3/2$ peak separating the two $c=1$ phases seems to emerge ~\cite{Rossotti_2017}.
Such a peak suggests  an emergent Ising  behavior at the transition, as the 2D Ising model at criticality carries an additional contribution $c=\frac{1}{2}$ on top of the background $c=1$ bosonic LL theory. Such conclusion is in-line with an effective field theory based on a two-fluid description valid in the CLL phase (called the strong paired phase in \cite{Kane2017} near by the transition where some pairs start dissociating. Such effective low-energy field theory has been shown to support an Ising phase transition  with  $c=\frac{1}{2}$ \cite{Kane2017}.

%The natural interpretation of \ps{the overall} scenario is the occurrence, beyond the critical point featuring an emergent $c=1/2$ Ising field at the transition, of the unconventional CLL we expect to arise from the corresponding from the strong-coupling arguments.
It is worth remarking that the numerical results are affected by a lack of convergence of the DMRG algorithm both at strong coupling and at the critical point. Thus, more accurate simulations with up to $2800$ kept states are actually necessary to show a proper scaling of the fitted value of the central charge compatible with a $c=\frac{3}{2}$ critical point and to provide clear evidence of a $c=1$ phase on the strong coupling side of the transition.

\begin{figure}[t]
\centering
\includegraphics[width=0.8\columnwidth]{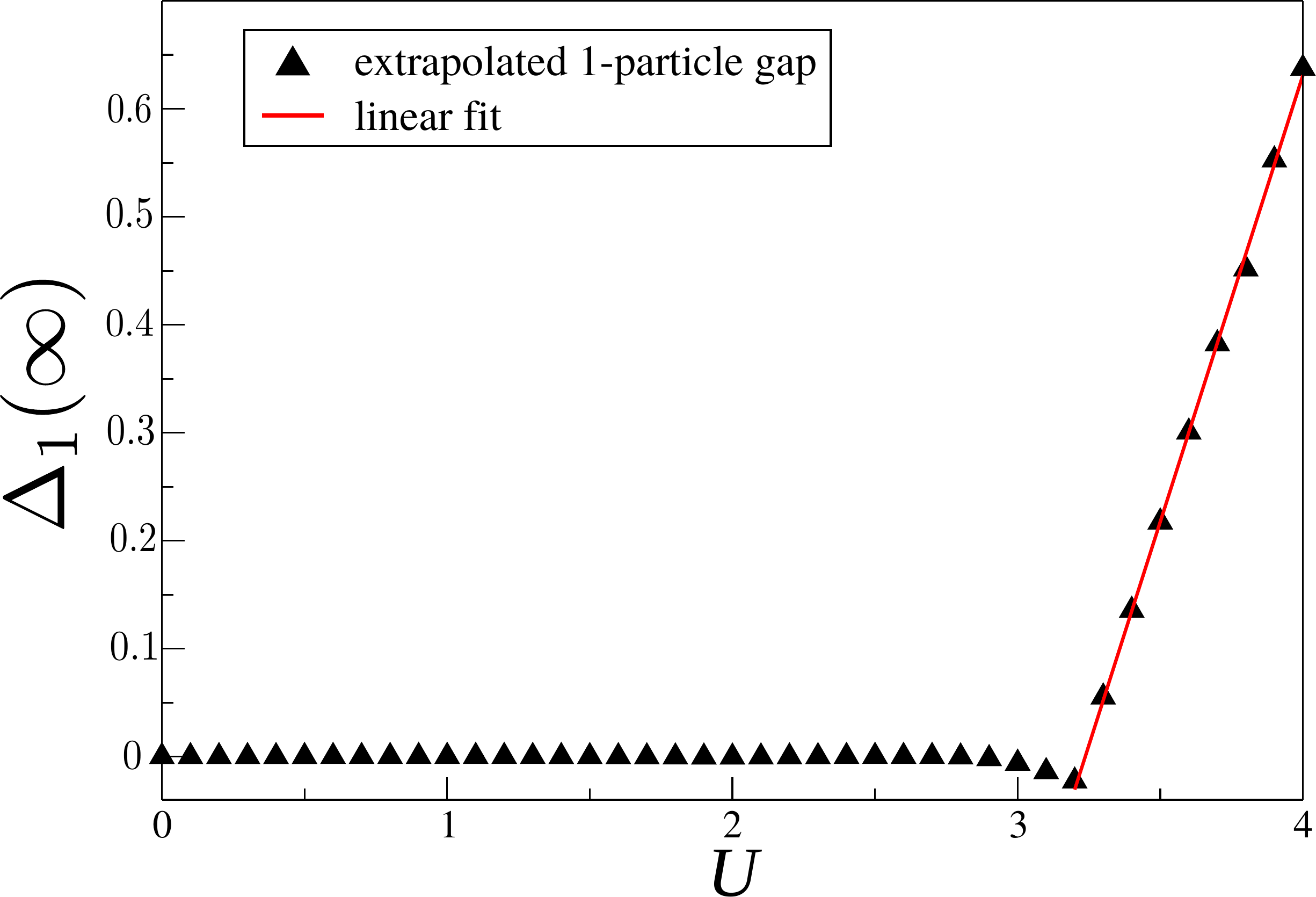}
\caption{Single-particle gap $\Delta_{1}(\infty)$ obtained by extrapolating  finite size gaps \eqref{gap} with $L=20,40,60,80$ on the same line as in Fig.~\ref{fig:free_energy_c_neg_02}. }
\label{fig:1gap_neg_02}
\end{figure}

In order to probe the transition to a liquid phase whose physical behaviour is exhaustively captured by pairing signatures, we investigate the spectral properties of the system by computing the single-particle gap and the pair-gap across the critical point. We observe in Fig.~\ref{fig:1gap_neg_02} the opening of a finite single-particle gap accompanied with a vanishing pairing gap, which in turn confirms the gapless nature of the $c=1$ CLL phase beyond the critical point. Notice that the opening of the single-particle gap agrees well with a linear behavior expected for the Ising universality class.

\begin{figure}[t]
\centering
\includegraphics[width=0.85\columnwidth]{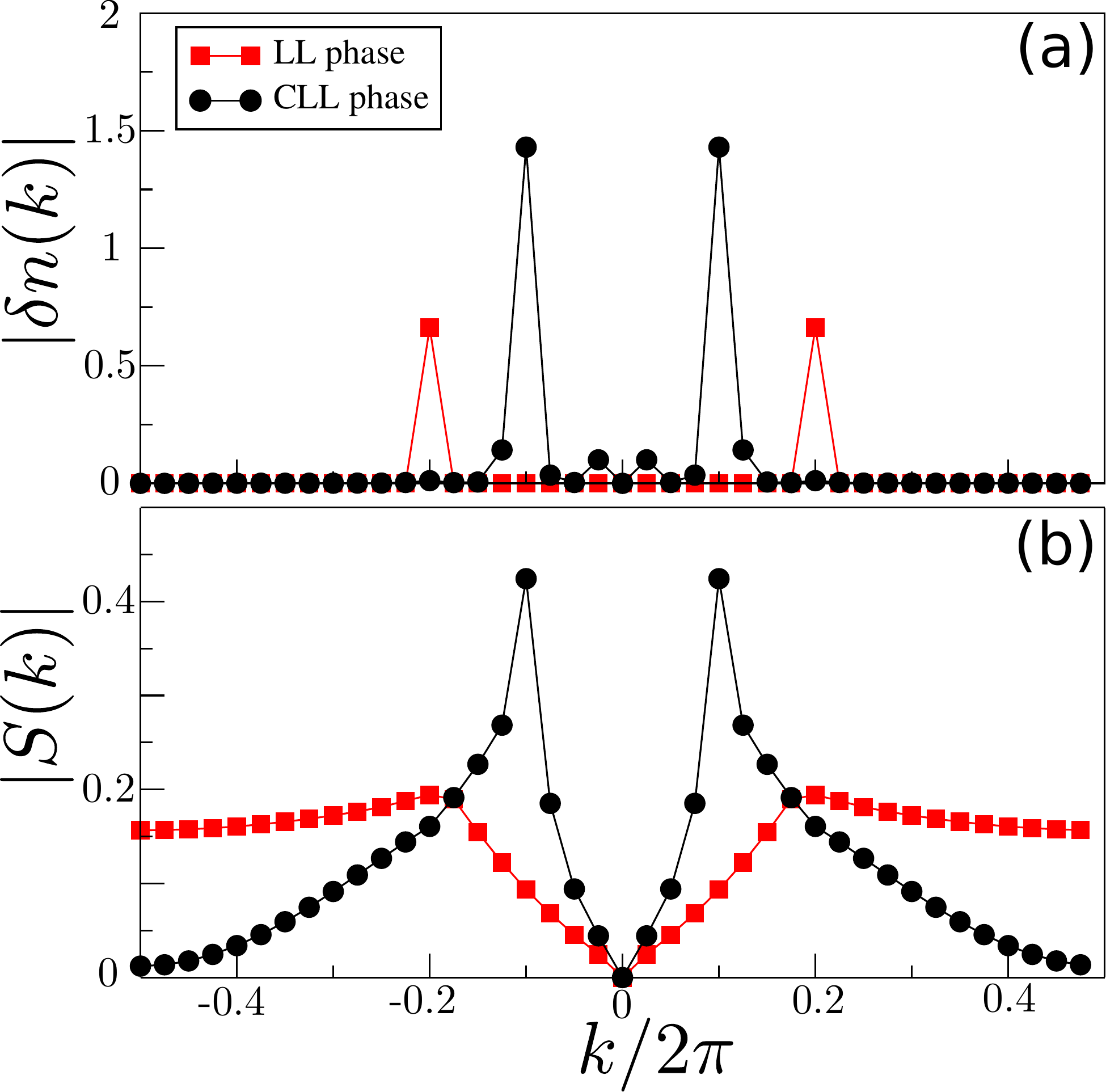}
\caption{\textsf{\textbf{(a)}} Density fluctuations Fourier transform \eqref{density_fourier}  and density structure factor \eqref{structure_factor} \textsf{\textbf{(b)}} on the same line as in Fig.~\ref{fig:free_energy_c_neg_02} for $L=40$. $U=1$ for the LL phase and $U=5$ for the CLL phase.}
\label{fig:nk_Sk_neg_02}
\end{figure}

%\subsubsection{Characterization of CLL phase}

To fully characterize such a novel state of matter, we investigate the behaviour of the Fourier transform of the density profile $\delta n(k)$  and of the density structure factor $S(k)$.  The reason for such a choice lies in the bosonization prediction that the expectation value of the aforementioned observables is given by an expansion whose lowest order harmonics oscillate with wavevector $k=2\pi \rho$, $\rho$ being the mean density of the microscopic granularity of the Luttinger liquid phase. More explicitly, the lowest order contributions to the density-density correlations read~\cite{giamarchi_quantum_2010}:
\begin{equation}
\langle\rho (x) \rho(0)\rangle = \frac{A}{x^{2}}+B\frac{\cos\left(2\pi \rho x\right)}{x^{2K}},
\end{equation}
where $A$ and $B$ are non-universal amplitudes. For the LL phase, we have $\rho = n$ while for the CLL phase, we expect $\rho = n/2$. 
The two phases are thus signaled by their corresponding peaks in both $\delta n(k)$ and $S(k)$ at wave-vectors $k = 2\pi \rho$.
As shown in Fig.~\ref{fig:nk_Sk_neg_02}(a-b), we do observe a shift in the momentum peak from $k=2\pi\cdot \frac{1}{5}$ to $k=2\pi\cdot \frac{1}{10}$, indicating the emergence of pairs as the elementary constituent of the CLL phase.

\subsection{Repulsive regime}

\begin{figure}[t]
\centering
\includegraphics[width=\columnwidth]{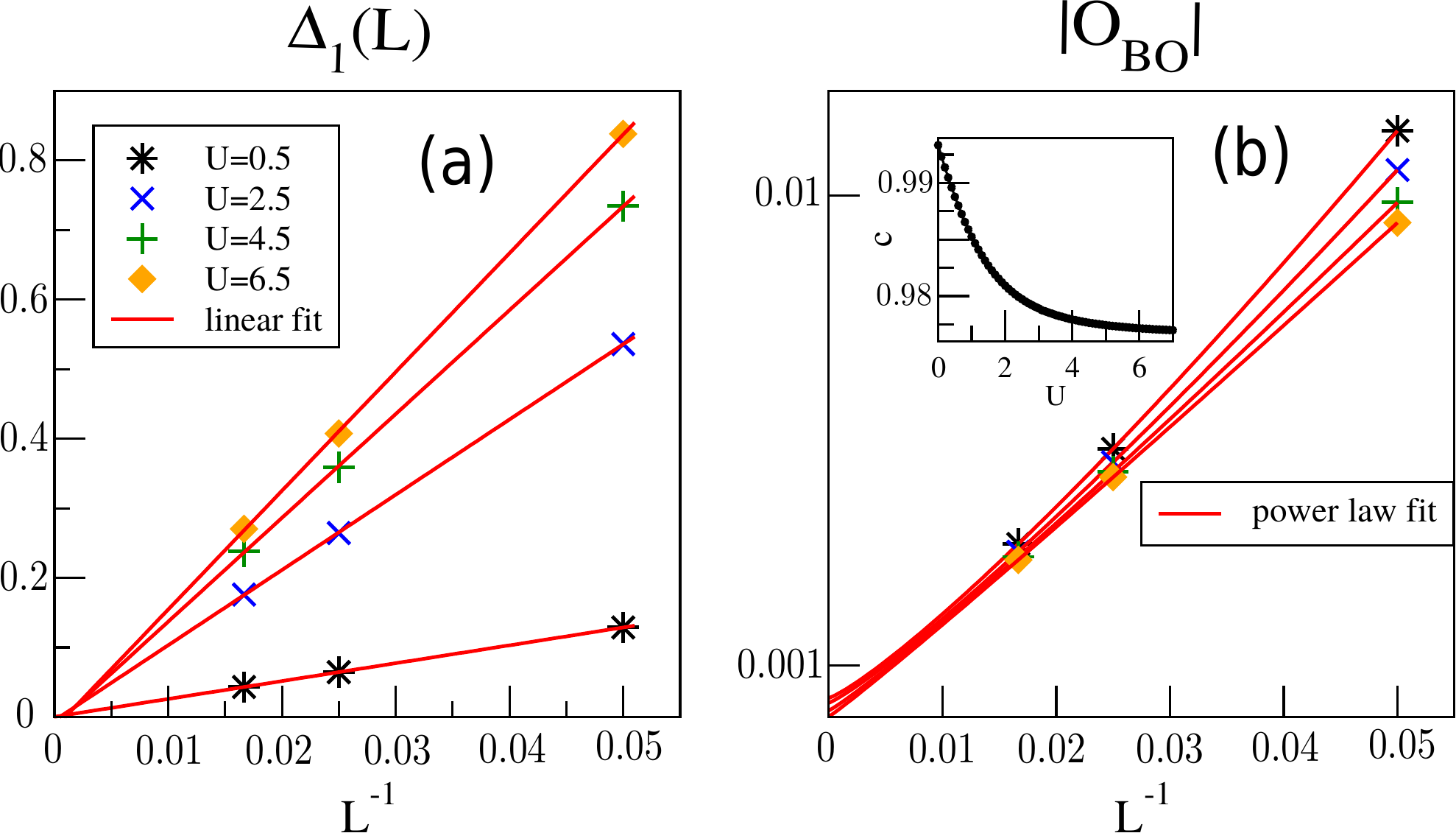}
\caption{DMRG results for $n=\frac{1}{5}$ along the $U=U_{2}=U_{1}$ line.
\textsf{\textbf{(a)}} finite size scaling of the single-particle gaps \eqref{gap}. \textsf{\textbf{(b)}} finite size scaling of the BO parameter \eqref{bond_order} using a power law $a_{0}+a_{1}L^{-a_{2}}$.
\textit{Inset}: extrapolated central charge from systems with $L=60,80,100,120,140$. }
\label{fig:1gap_BO_pos_02}
\end{figure}

From the study of the classical limit, the system is expected to behave as a regular LL for repulsive interactions, without exhibiting any transition to some alternative phases.
In order to support such a claim, we show in Fig.~\ref{fig:1gap_BO_pos_02}(a) the finite-size scaling for single-particle gap, which manifestly scales to zero as a function of the inverse system size $L^{-1}$ over the whole sampled region up to comparatively large interaction strength.
Furthermore, the finite-size entropy in Fig.~\ref{fig:PDn02} does not show any harbingers of phase transitions for the whole repulsive region. 
Looking at the extrapolated central charge for a wide range of values along the line $U_{2}=U_{1}$ shows deviation from $c=1$ value below $2.3\%$, which may be ascribed to the effect of interactions in a finite size and perfectly compatible with a repulsive LL. 
% {\color{red}[L.M. Shall we put the red line $U_1=U_2$ in the phase diagram?]}
As a conclusive remark, the finite size scaling of the BO parameter is presented in Fig.~\eqref{fig:1gap_BO_pos_02}, where it is shown to extrapolate to zero in the infinite size limit over the whole range of sampled values.  

\section{Phase diagram for $\mathbf{n= \frac 13}$} \label{third}

\begin{figure}[b]
\centering
\includegraphics[width=0.9\columnwidth,clip]{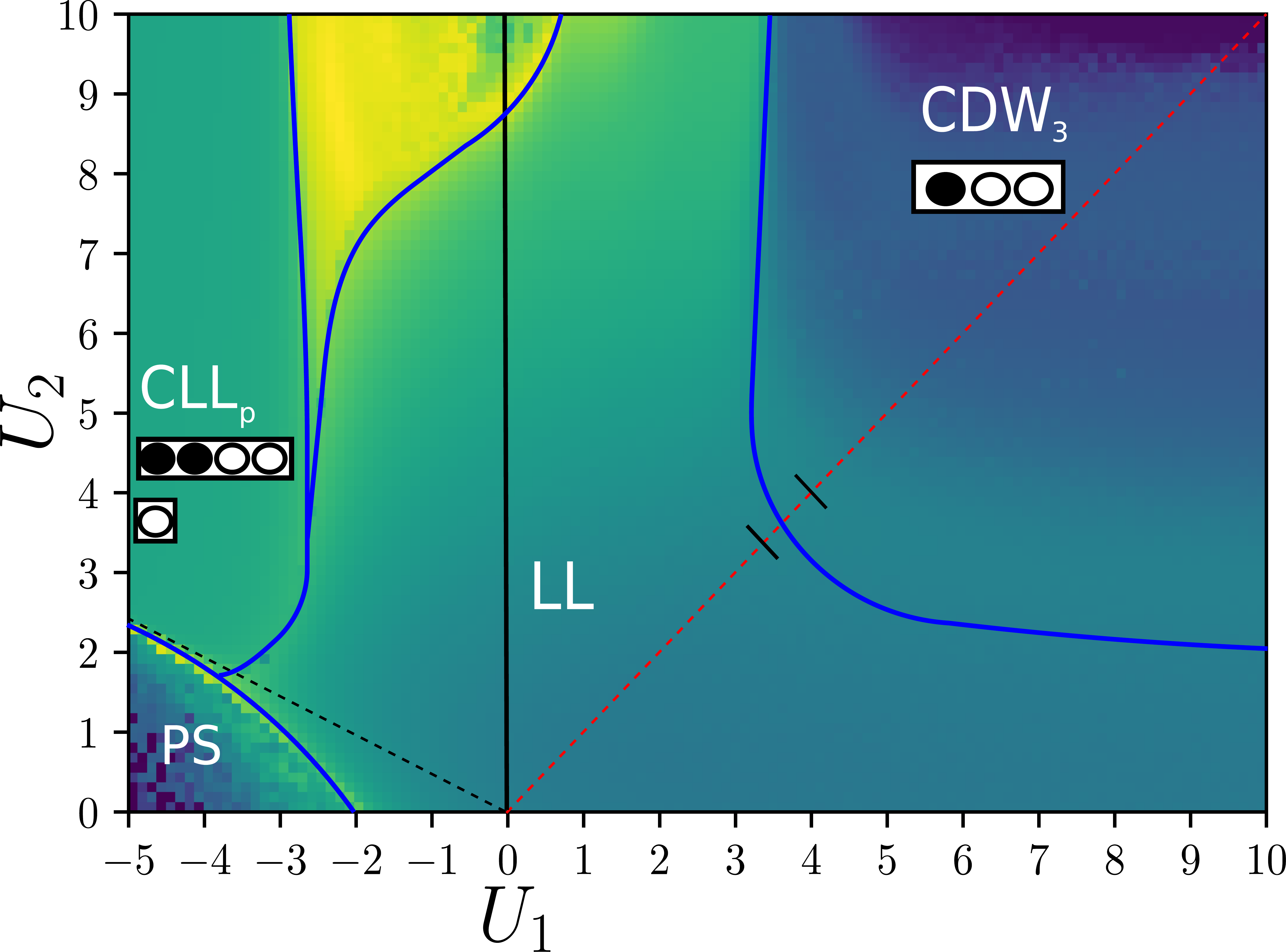}
\caption{Phase diagram for $n=\frac{1}{3}$. Color map for the background displays the entanglement entropy on a $L=42$ chain with PBC. Black lines are classical transition lines obtained neglecting quantum fluctuation. Additional numerical simulations are presented for the points lying on the red lines. Blue lines are a guide to the eye for the main phase boundaries.}
\label{fig:PDn03}
\end{figure}

The phase diagram depicted in Fig.~\ref{fig:PDn03} for $n = \frac 13$ is qualitatively very similar to that in Fig.~\ref{fig:PDn02} for $n=\frac 15$ in what concerns the LL, CLL and PS phases. The main differences are (i) the appearance of a highly entropic phase for $|U_1| \ll1$ and $U_2 \gg 1$; and (ii) the emergence of a gapped insulating phase that appears exclusively for $n= \frac 13$. 
This section is mainly devoted to the characterization of this latter phase, which displays CDW order with one particle every three sites, named CDW$_3$.
It emerges at strong-coupling in the purely repulsive regime, as a result of the commensurability effect arising from the interplay between the density and the interaction range. 

The discussion of the nature of the large $U_{2}$ highly entropic phase highlighted above is postponed to section (\ref{fourth}), where we will provide an analytical treatment of the latter showing that it appears for $\frac 13\leq n < \frac 12$ and agreement with the related numerical data.

\subsection{Classical limit}

For $n= \frac 13$, the qualitative features of the classical-limit configurations remain essentially unchanged for $U_{1}<0$. 
On the other hand, the classical configurations in the $U_{1}>0$ regime are obtained from the periodic repetition of a block $C$ with unit cell ($\bullet\circ\circ$), since it realizes the lowest classical energy density $\epsilon_{GS}=0$. 
Such a result is consistent with the general statement that a fermionic system with repulsive soft-shoulder interactions of range $R$ stabilizes, in the classical limit and at filling $n=\frac{1}{R+1}$, a periodic arrangement of period $R+1$ with one particle each $R+1$ sites~\cite{szyniszewski_fermionic_2018}.
This periodic arrangement is optimal in the classical limit independently of the ratio $\frac{U_{2}}{U_{1}}$ and the quantum phase expected to naturally arise at sufficiently strong coupling is then the CDW$_3$.
Then, a critical line must exist in the phase diagram of Fig.~\ref{fig:PDn03}
separating the weak-coupling LL regime from this CDW$_3$ phase.

\subsection{The CDW$_3$ insulator}

\begin{figure}[t]
\centering
\includegraphics[width=\columnwidth]{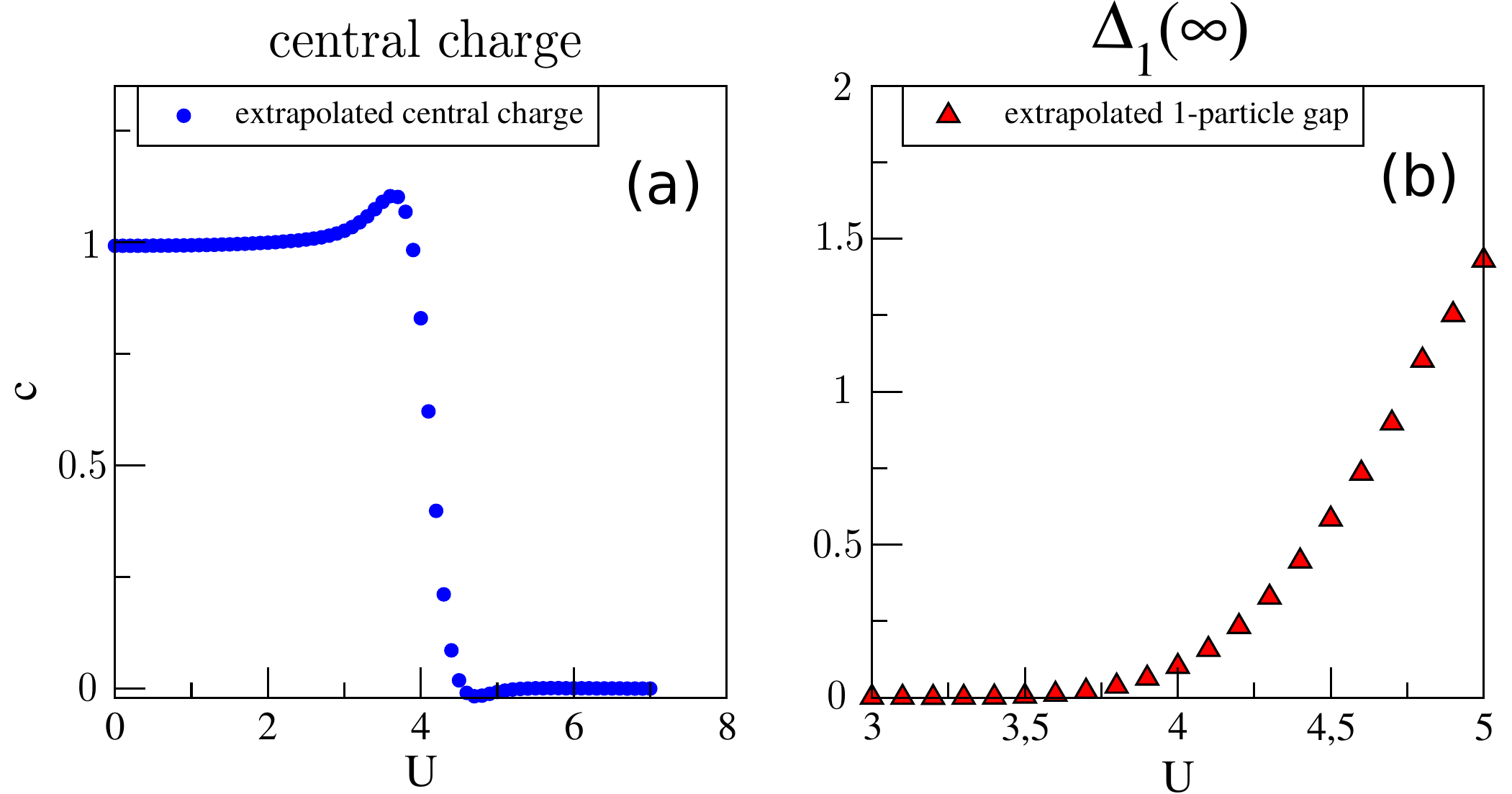}
\caption{DMRG results for $n=\frac{1}{3}$ along the $U=U_{2}=U_{1}$ line.  \textsf{\textbf{(a)}} extrapolated central charge from $L=61,82,100,121,142$. 
\textsf{\textbf{(b)}} extrapolated single-particle gap from sizes $L=73,97,121$ showing its opening around the critical point.}
\label{fig:c_gap_13}
\end{figure} 

In order to study the onset of the CDW$_3$ phase along the line $U_{2}=U_{1}$, we compute the single-particle gap $\Delta_{1}$, the central charge $c$, the behaviour of single-particle and pair correlation functions, and the Fourier spectrum of the density profiles. 
In Fig.~\ref{fig:c_gap_13}(a), the extrapolated central charge, obtained from fitting the entanglement profile and then extrapolating the finite-size results with some polynomial fit, nicely displays a step from 1 to 0 as expected when entering a gapped phase.
At the same time, the opening of the single-particle gap is shown in Fig.~\ref{fig:c_gap_13}(b) after extrapolating the finite-size gaps with a polynomial law. The qualitative features of Fig.~\ref{fig:PDn03} support the fact that the critical line should reach $U_2=0$ when $U_1/t \to \infty$, since any finite $U_2$ should stabilize a weak CDW$_3$ and similarly along the $U_2>0$ axis. When $U_1=0$, creating pairs does not cost anything so other classical configurations compete with the CDW$_3$ one.

\begin{figure}[t]
\centering
\includegraphics[width=\columnwidth]{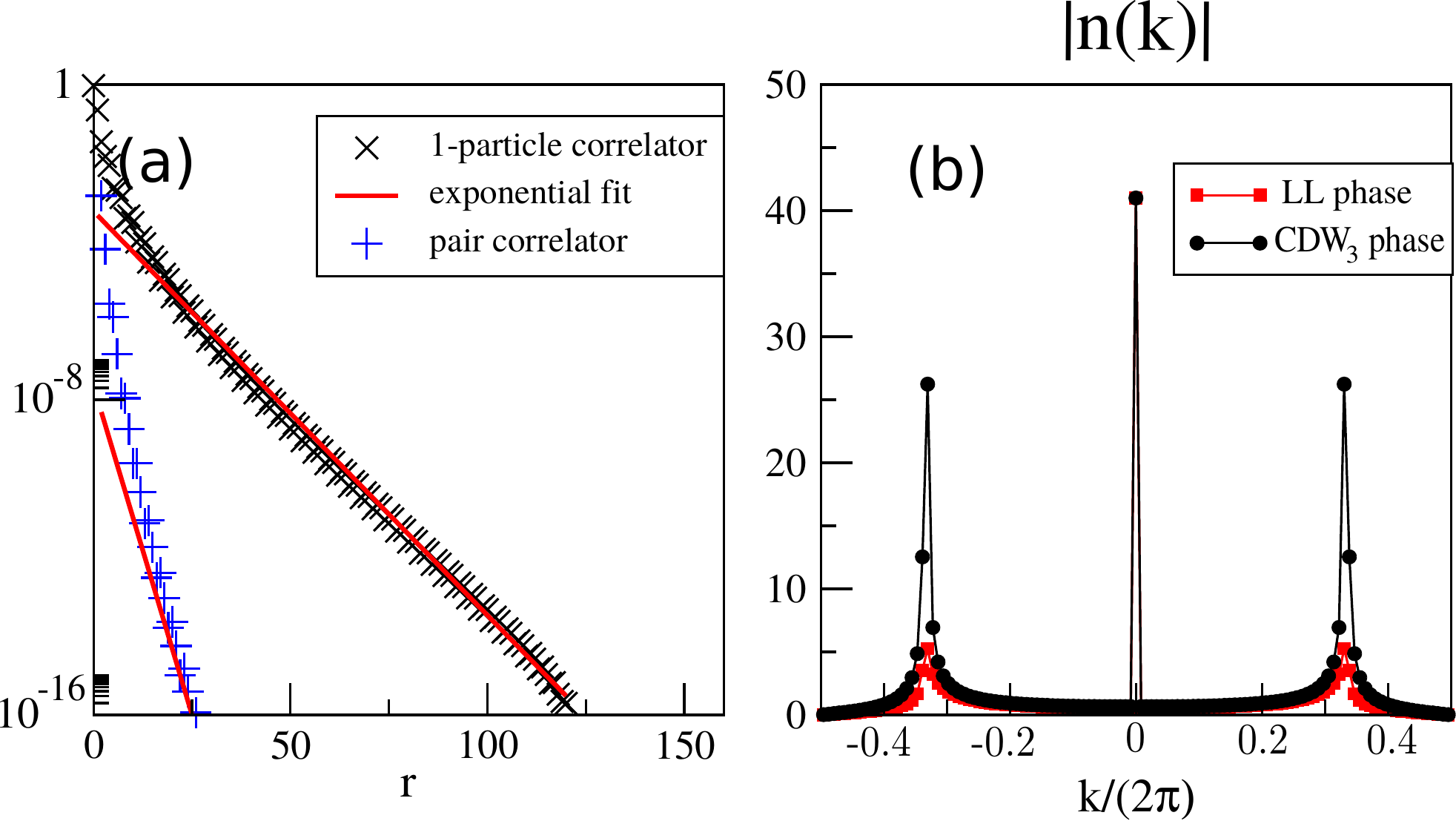}
\caption{DMRG results for $\frac{1}{3}$ along the $U=U_{2}=U_{1}$ line. \textsf{\textbf{(a)}}
 decay of the single-particle and pair correlators $G(r)$ and $P(r)$ as a function of separation for $L=121$ and $U=5$, deep in the CDW phase.
\textsf{\textbf{b)}}: density profile Fourier spectrum $n(k)=\delta n(k)+n\delta_{k,0}$ for $U=3$ (LL phase) and $U=5$ (CDW$_3$ phase).}
\label{fig:corr_nk_13}
\end{figure}
 
Since the finite energy gap in the excitation spectrum is naturally accompanied with the emergence of a finite correlation length, we monitor in Fig.~\ref{fig:corr_nk_13}(a) the decay law of the single-particle and pair correlators deep in the massive CDW$_3$ phase.
The outcome of the numerical simulations confirms an exponential decay, in  agreement with the phenomenology of gapped insulating phases.

Finally, the Fourier spectrum of the density profile is displayed in Fig.~\ref{fig:corr_nk_13}(b), so that the enhancement of the Fourier component associated to the quasi-momentum $k=2\pi\cdot\frac{1}{3}$ of the underlying crystal-like arrangement of the particles is observed.
The height of the related peak, being of the same order of magnitude as the extensive zero mode, strongly suggests the onset of long-range crystalline order, as opposed to what is observed in the LL phase.
We expect the transition between the LL regime and the CDW$_3$ region
to be of the Berezinsky-Kosterlitz-Thouless type~\cite{giamarchi_quantum_2010}, with an expected critical Luttinger exponent of $K_c=2/9$ at the transition line.
Interestingly, although the density correlations may not decay slow enough in our case, we notice that our model, or deformation of it, may be suitable for stabilizing an incommensurate floating phase~\cite{Bak1982} in the vicinity of the CDW$_3$ phase. It would be a region were the LL phase has density-correlations decaying so slow that the structure factor diverges at some incommensurate wave-vectors. Recent intensive numerical works on qualitatively related models have shown the possible realization of such unusual behavior~\cite{Chepiga2019,Rader2019}.

\section{Phase diagram for $\mathbf{n= \frac 25}$} \label{fourth}

\begin{figure}[b]
\centering
\includegraphics[width=\columnwidth,clip]{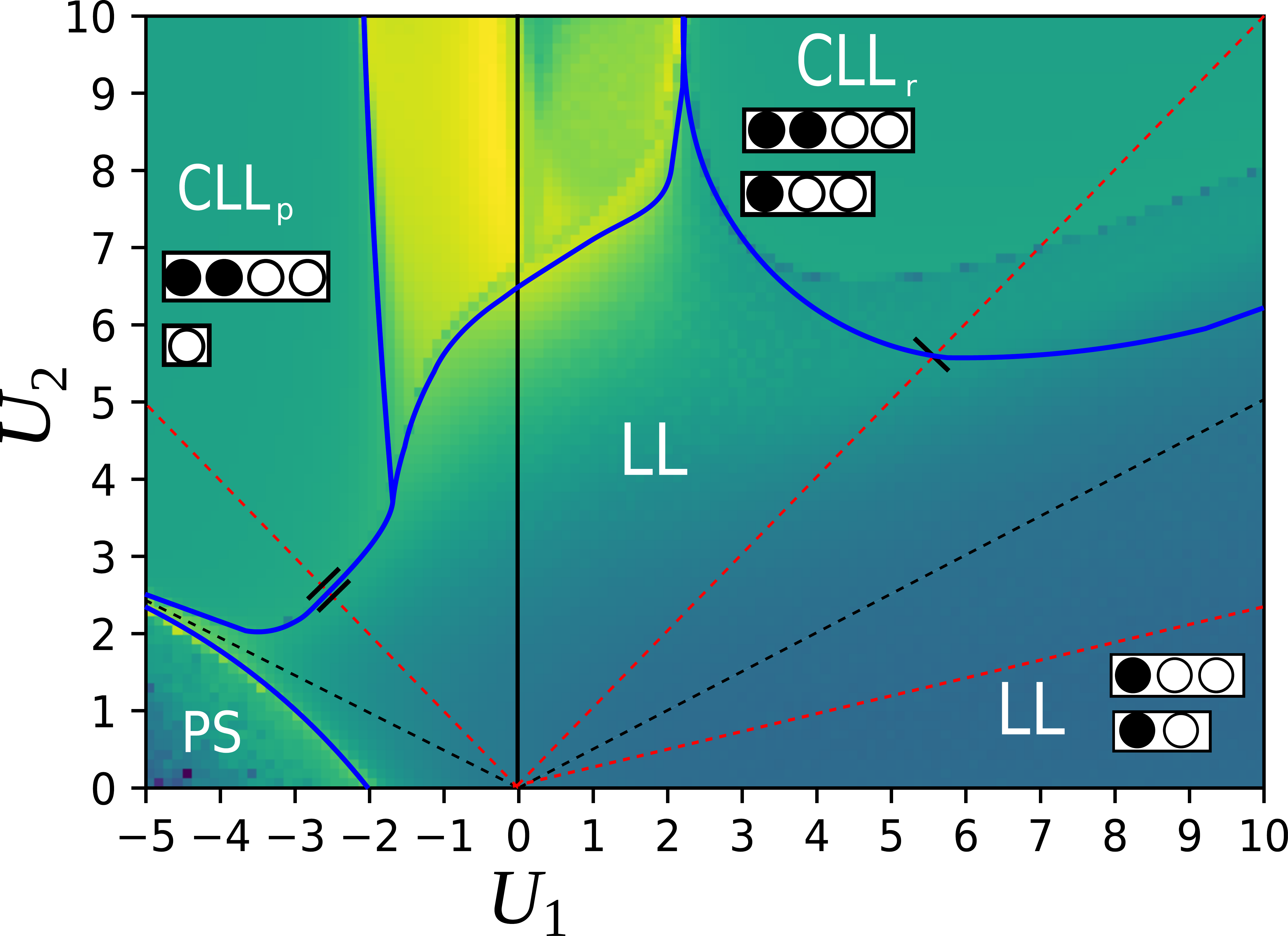}
\caption{Phase diagram for $n=\frac{2}{5}$. Colormap for the background display the entanglement entropy on a system with $L=30$ and PBC. Black lines are classical transition lines obtained neglecting quantum fluctuation. Additional numerical simulations are presented for the points lying on the red lines. Blue lines are a guide to the eye for the main phase boundaries.}
\label{fig:PDn04}
\end{figure}

The two goals of the present section consist in showing that in the range of densities $\frac 13 < n < \frac 12$ (i) the pair CLL phase in the attractive regime  survives, and (ii) in the repulsive region a novel and frustration-induced CLL phase appears.
The two paired phases have incompatible signatures; whereas the former result points out the robustness of the qualitative features of the phase diagram for $U_1<0$, the second results shows that the region $U_1>0$ has a complete restructure in this density regime. Our study focuses on $n = \frac 25$ and a phase diagram drawn from entropic considerations is in Fig.~\ref{fig:PDn04}.

% at higher density, thereby confirming the robustness of the qualitative features of the phase diagram for $U_{1}<0$ and, second, providing evidence for the appearance of a frustration-induced CLL phase in the repulsive region, featuring incompatible signatures with its attractive counterpart. 

\subsection{Classical limit}

The study of the region $U_1<0$ of the phase diagram in the classical limit $t=0$ outcomes the same features obtained for lower densities, name a critical line at $U_2 = - U_1/2$ separating PS from CLL.

% The chosen filling is typical of the density range $1/3\leq n \leq 1/2$ and shows a very similar behavior in the negative $U_1$ region as the previous diagrams with lower fillings, controlled by phase separation and the same classical limit analysis.
% In particular, we will give in the next subsection numerical evidence for the survival of the $U_{1}<0$ CLL phase at $n=0.4$.

Instead, the classical-limit analysis of the phase diagram is richer in the repulsive region $U_1>0$, and depends on the value of the ratio $U_{2} / U_{1}$. 
Whenever $U_{2} / U_{1} < 1/2$, the generic classical ground state configuration is given by any permutation of blocks $C$ with blocks $D$ ($\bullet\circ$) satisfying $N_{C}=L-2N,N_{D}=3N-L$. Since the elementary degrees of freedom in classical limit are single fermions and the effective strong-coupling description of the system is given once again by Eq.~\eqref{XX}, a naive hypothesis for the collective behaviour of the system is the survival at all couplings of a standard LL phase.~\cite{mattioli_cluster_2013,dalmonte_cluster_2015}

Conversely, if $U_{2} / U_{1} > 1/2$, the resulting classical ground states are generated by all possible permutations of blocks $C$ with blocks $A$, under the constraints $N_{C}=L-2N$, $N_{A} = (3N-L)/{2}$. The strong-coupling dynamics turns out to be described again by a XX-Hamiltonian.
This demonstrates that any strong-coupling phase is adiabatically connected to a $c=1$ conformal gapless phase. On the other hand, in contrast to the classical ground state structure below the critical line, the fundamental blocks in the present case realize a nontrivial microscopic structure consisting of a mixture of single particles and frustration-induced pairs. The latter is to be interpreted as an ideal classical platform for the emergence of exotic CLL phases in the $t\neq 0$ regime.~\cite{mattioli_cluster_2013,dalmonte_cluster_2015}
Indeed, the fundamental granularity of the expected liquid behaviour will differ from the bare fermionic particles in terms of which the model is defined, giving rise to characteristic signatures that will be illustrated with numerical results.

As a last step, the classical limit of the system along the $U_{1}=0$ axis deserves a special treatment. In such a case, the fundamental blocks whose permutations generate the whole degenerate ground state manifold are the blocks of type $A\,(\bullet\bullet\circ\circ),\,B\,(\circ),\,C\,(\bullet\circ\circ)$. Thus, the ground state manifold includes, among others, the degenerate classical ground states presented for the attractive and repulsive regimes. This opens new scenarios for the physics of the system. Imposing the standard filling constraints:
\begin{align}\label{phase_sep_cl_lim}
\begin{cases}
2N_{A}+N_{C}=N, \\
4N_{A}+3N_{C}+N_{B}=L
\end{cases}
\end{align} 
one obtains, in the representative case $n=\frac{2}{5}$, the expressions $N_{A}=\frac{L}{10}+\frac{N_{B}}{2},\, N_{C}=\frac{L}{5}-N_{B}$, i.e., the generators of the ground state subspace are parameterized by the value of $N_{B}$, which is a free parameter interpolating between the attractive regime ground state manifold ($N_{B}=\frac{L}{5}$) and its repulsive regime counterpart ($N_{B}=0$).

In order to shed light on the phase emerging from the above classical limit once quantum fluctuations are introduced, we compute the strong coupling effective Hamiltonian to first order:
\begin{equation} \label{phase_sep_strong_coupling}
H\approx -t\sum_{j}\left(\hat{M}^{\dag}_{j}\hat{M}_{j+1}+h.c.\right)
\end{equation} 
where $\hat{M}_{j}=|{B}\rangle_{j}\langle{C}|_{j}$. Thus, from the physical point of view, the blocks $B$ and $C$ obey an effective spin-$\frac{1}{2}$ XX dynamics, whereas blocks $A$ are completely immobile within the first order approximation. 

Let us now discuss the density dependence of the collective behaviour of the system in such limit. First, we observe that the spin-$\frac{1}{2}$ XX Hamiltonian governing the dynamics of blocks $B$ and $C$ favors energetically states exhibiting hybridization between blocks $B$ and $C$. Hence, in order to increase their kinetic energy, the ground state of the system tends to minimize $N_{A}$. There are then two situations. If $ n<\frac{1}{3}$, one can set $N_{A}=0$ in equations \eqref{phase_sep_cl_lim}, leading to $N_{B}=L-3N_{C}$ and $N_{C}=N$ and a regular LL phase as observed in Fig.~\ref{fig:PDn02}.

When $\frac{1}{3}\leq n<\frac{1}{2}$, one must have $N_{A}\neq 0$.
For blocks $B$ and $C$, as the ground-state of the spin-$\frac{1}{2}$ XX Hamiltonian is in the zero magnetization sector, we assume that the optimal condition is $N_{B}=N_{C}$. Specializing to the case $n=\frac{2}{5}$, we get 
$N_{B}=N_{C}=\frac{L}{10}$ and $N_{A}=\frac{3}{20}L$.
As a consequence, we conjecture that the system enters a new regime of phase separation, where the CDW$_2$ phase coexists with a LL region at effective density $\frac{N_{C}}{3N_{C}+N_{B}}=\frac{1}{4}$. Our claim is supported by the results of Fig.~\ref{fig:PDn03} and Fig.~\ref{fig:PDn04}, where the emergence of a phase separated region in the limit $\frac{U_{2}}{|U_{1}|}\gg 1$ is manifested by a larger entropy. This observation stems from the presence of the LL phase in the middle of the chain, as discussed in Sec.~\ref{yellow_region_subsection}.

\subsection{Numerics: attractive regime}

\begin{figure}[t]
\centering
\includegraphics[width=\columnwidth]{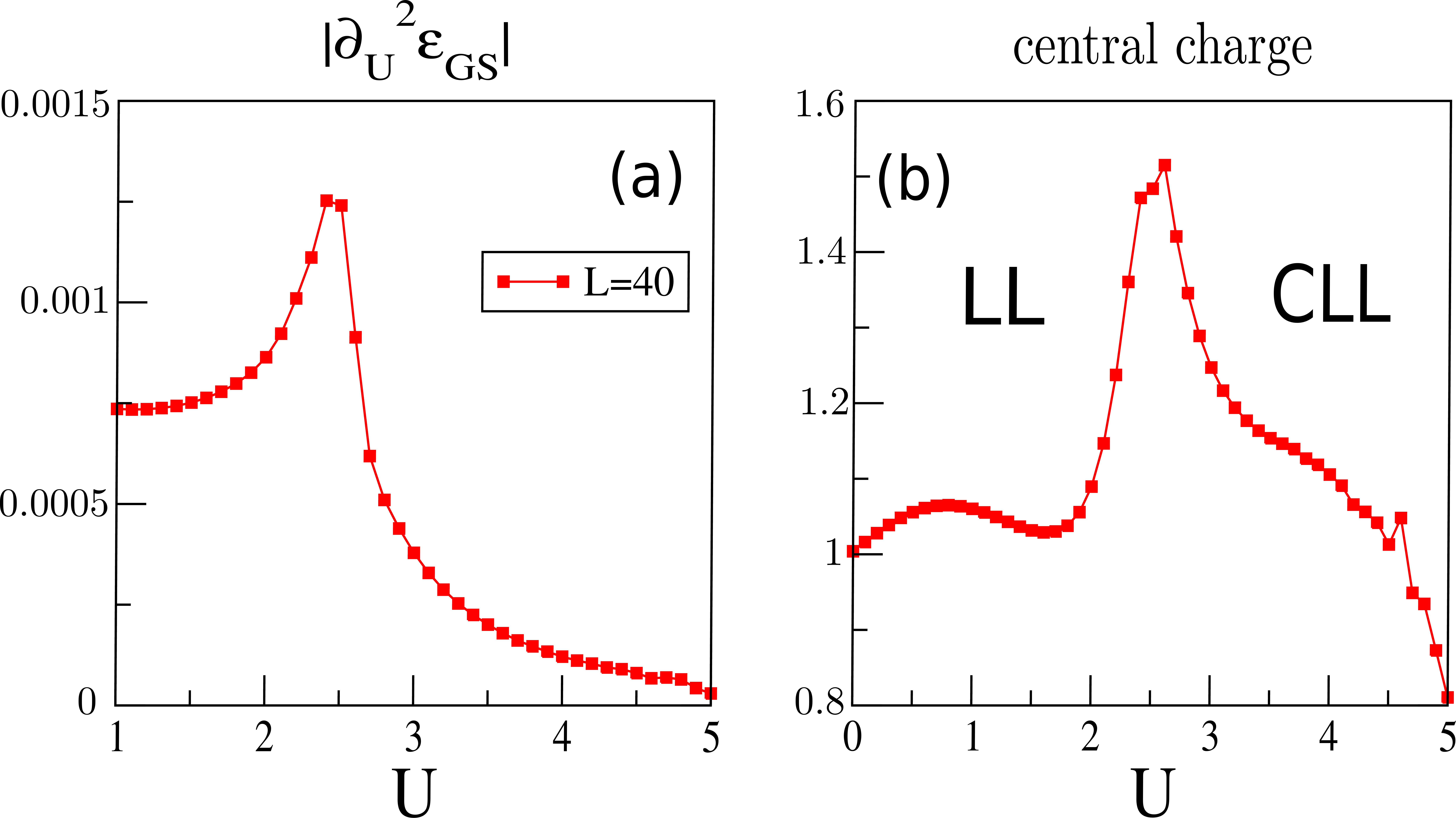}
\caption{DMRG results for $n=\frac{2}{5}$ along the $U=U_{2}=-U_{1}$ line for $L=40$ and PBC. \textbf{\textsf{(a)}} Ground state energy density curvature \eqref{curvature} The shape of the profile is likely to be a non-analytic function of the control parameter $U$, locating the critical point within $-2.5 \leq U_{1} \leq -2.4$.
\textbf{\textsf{(b)}} Extrapolated central charge from \eqref{cardy-calabrese}. Data once again support a $c=\frac{3}{2}$ critical point, where an additional Ising degree of freedom emerges on top of the bosonic $c=1$ contribution, suggesting the universality class of the 2D Ising model.}
\label{fig:c_free_energy_neg_04}
\end{figure}

\subsubsection{Characterization of the transition}

Keeping in mind the classical limit picture of the system for $n=\frac{2}{5}$ and $U_{2}>-U_{1}/2$ in the attractive region, we expect a transition from the weak-coupling LL phase to an unconventional CLL phase with pairing. 
Thus, we investigate the characteristic signatures of the critical point separating the two phases by computing relevant observables along the cut $U = U_{2}=-U_{1}$.
The first quantity we monitor is the second derivative of the ground state energy density, whose characteristic non-analyticity presented in Fig.~\ref{fig:c_free_energy_neg_04}(a) is analogous to that of Fig.~\ref{fig:free_energy_c_neg_02}(a) for $n=\frac{1}{5}$, which points towards the presence of a critical point.
In addition, the second quantity of interest is the central charge $c$  obtained by fitting the entropy profiles to formula \eqref{cardy-calabrese}. The result, shown in Fig.~\ref{fig:c_free_energy_neg_04}(b) is compatible with the presence of two $c=1$ phases separated by an exotic critical point with central charge $c = 3/2$, suggesting the emergence of a critical Ising degree of freedom at the transition.

\begin{figure}[t]
\centering
\includegraphics[width=0.85\columnwidth,height=6.5 cm]{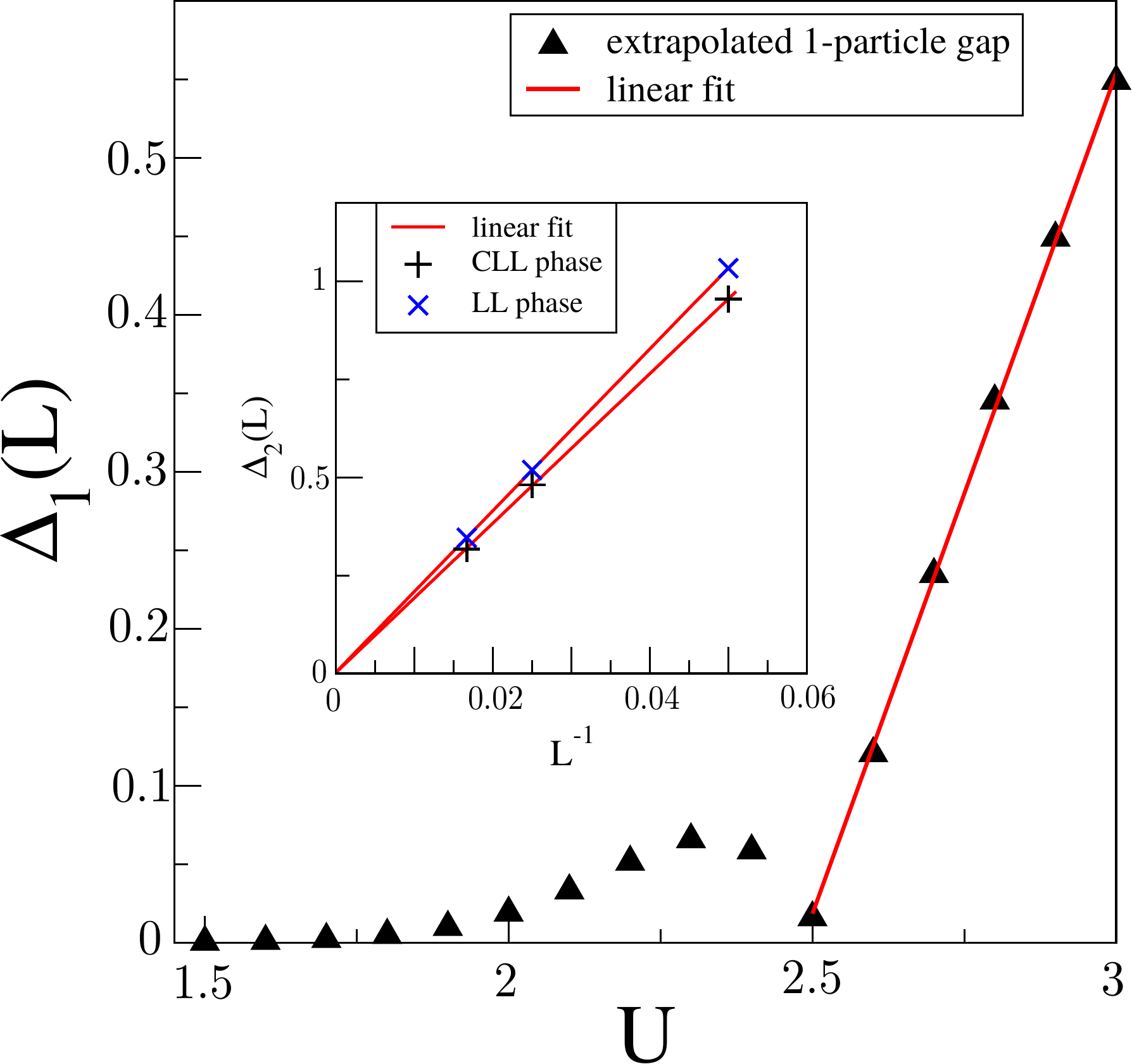} 
\caption{DMRG results for $n=\frac{2}{5}$ along the $U=U_{2}=-U_{1}$. Extrapolations of the single-particle gap $\Delta_{1}(N,L)$  across the transition between the two liquid phases. Finite size results are computed according from \eqref{gap} for $L=20,40,60$ and fitted to a linear model in $L^{-1}$. The orange line is a linear fit of the opening of the gap, consistent with the $2D$ Ising universality class.
\textit{Inset:} finite size scaling of the pair gap $\Delta_{2}(N,L)$  from \eqref{gap}, both in the LL phase ($U=1.5$) and in the CLL phase ($U=3.5$). They both scale to zero. }
\label{fig:1gap_2gap_neg_04}
\end{figure}

We now turn our attention to the low energy excitations of the model in the two phases. The results obtained by means of numerical simulations demonstrate clearly the opening of the single-particle gap at the transition, as shown in Fig.~\ref{fig:1gap_2gap_neg_04}, where the extrapolated value of the single-particle gap is plotted as a function of $U=U_{2}=-U_{1}$: a finite gap $\Delta_{1}(\infty)$ opens at the transition from the LL phase to the CLL phase. Furthermore, its dependence is shown to be linear with the distance from the critical point, supporting consequently a critical behaviour in agreement with the $2D$ Ising universality class.
On the other hand, we present in Fig.~\ref{fig:1gap_2gap_neg_04} the pair-gap $\Delta_{2}$ in both the LL and CLL phases.
We observe a scaling to zero in both phases when $L \to \infty$, which rules out the hypothesis of CDW formation or other possible gapped phases by asserting the gapless nature of pair degrees of freedom. 

\subsubsection{Characterization of the CLL phase}

\begin{figure}[t]
\centering
\includegraphics[width=0.85\columnwidth]{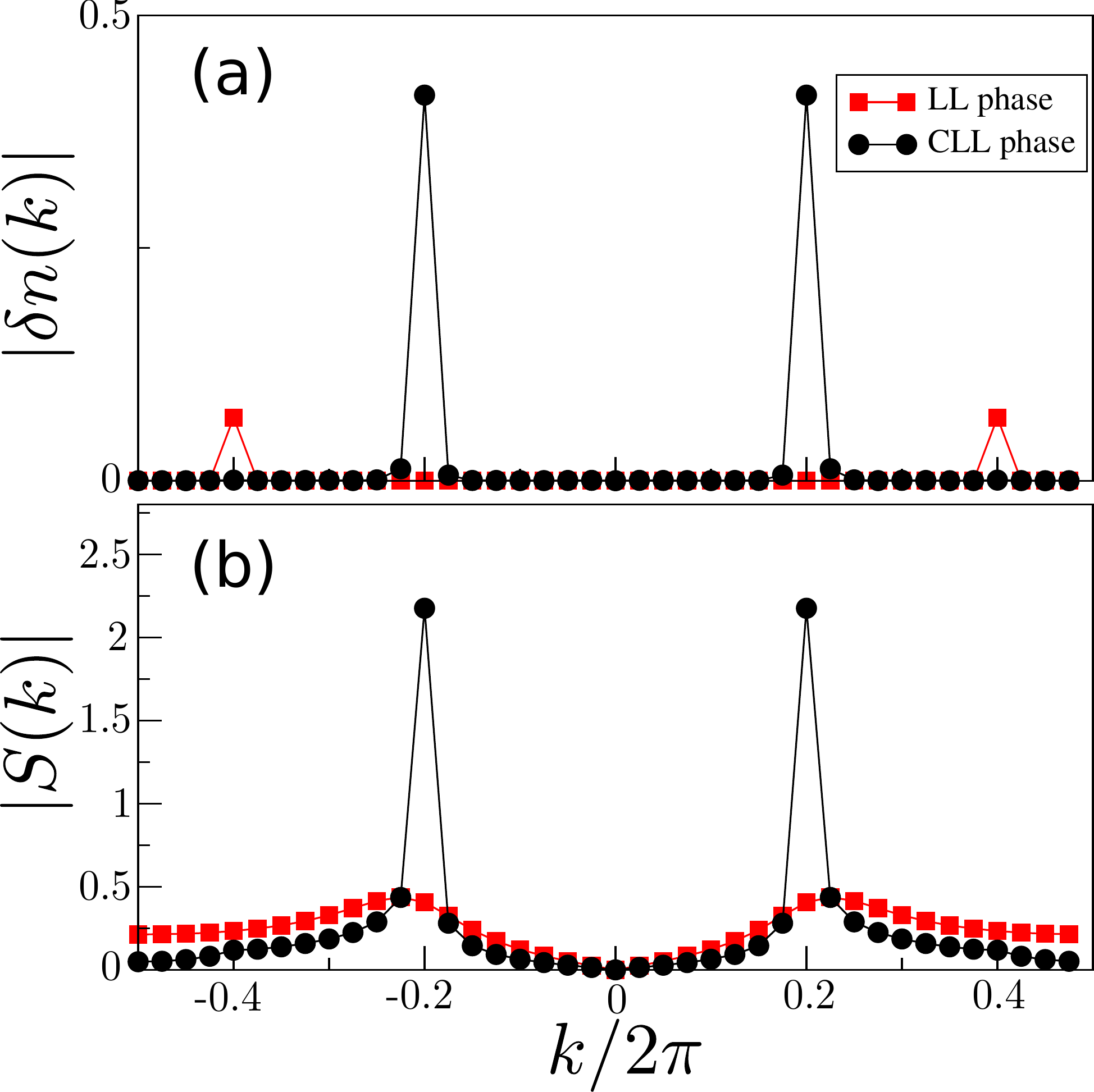}
\caption{DMRG results for $n=\frac{2}{5}$ along the $U=U_{2}=-U_{1}$ line for $L=40$ and PBC. \textbf{\textsf{(a)}} density fluctuations Fourier spectrum \eqref{density_fourier} and \textbf{\textsf{(b)}} density structure factor \eqref{structure_factor} both in the LL ($U=1.5$) and CLL phase ($U=3.5$). The momentum peak shift at value $k=2\pi\cdot \frac{1}{5}$ is incompatible with a standard LL theory and supports that the physics of the CLL phase in the attractive regime is ruled by pair degrees of freedom.}
\label{fig:nk_Sk_neg_04}
\end{figure}

Inspired by the procedure followed in the case $n=\frac{1}{5}$, we proceed with the analysis of the Fourier spectrum of density profiles and density structure factors. We therefore consider first the density profile in Fourier space on Fig.~\ref{fig:nk_Sk_neg_04}(a): while the leading peak is located at $k=2\pi \cdot \frac{2}{5}$ inside the LL phase region, the picture changes dramatically when considering the CLL phase, with a peak located at $k=2\pi\cdot \frac{1}{5}$, which is consistent with the above discussion and in striking disagreement with the conventional LL paradigm.

Similarly, the pronounced peak at the very same wave-vector observed in the static structure factor of the CLL ground state presented in Fig.~\ref{fig:nk_Sk_neg_04} is a further indication of the fundamentally different nature of the CLL phase, pointing decisively towards the effective deformation of the Bose surface that is not captured by LL theory.

\subsection{Numerics: repulsive regime}

\subsubsection{Characterization of the transition for $U_{2}>U_{1}/{2}$}

At density $n=\frac{2}{5}$, the nontrivial structure of the classical limit in the repulsive region for $U_{2}>{U_{1}}/ 2$ gives the opportunity of observing an exotic liquid phase.
Indeed, the main result of Refs.~[\onlinecite{mattioli_cluster_2013},\onlinecite{dalmonte_cluster_2015}] in the present setting is based on the observation that the fitted central charge profile along the line $U_{1}=U_{2}$ points towards the presence of a $c=3/2$ critical point, thus suggesting the same phenomenology as the one encountered in the attractive region. Furthermore, the behaviour of the single-particle and pair gaps has been shown in Refs.~[\onlinecite{mattioli_cluster_2013},\onlinecite{dalmonte_cluster_2015}] to coincide with the ones shown in the case of the negative $U_1$ CLL phase, suggesting that the related phase is of the same nature of the one discussed above. Conversely, its specificity with respect to the latter, whose origin may be traced back to its unique cluster structure, emerges when considering the spectral properties, as thoroughly discussed below.

\subsubsection{Characterization of the CLL phase for $U_{2} > U_{1}/{2}$}

\begin{figure}[t]
\centering
\includegraphics[width=0.85\columnwidth]{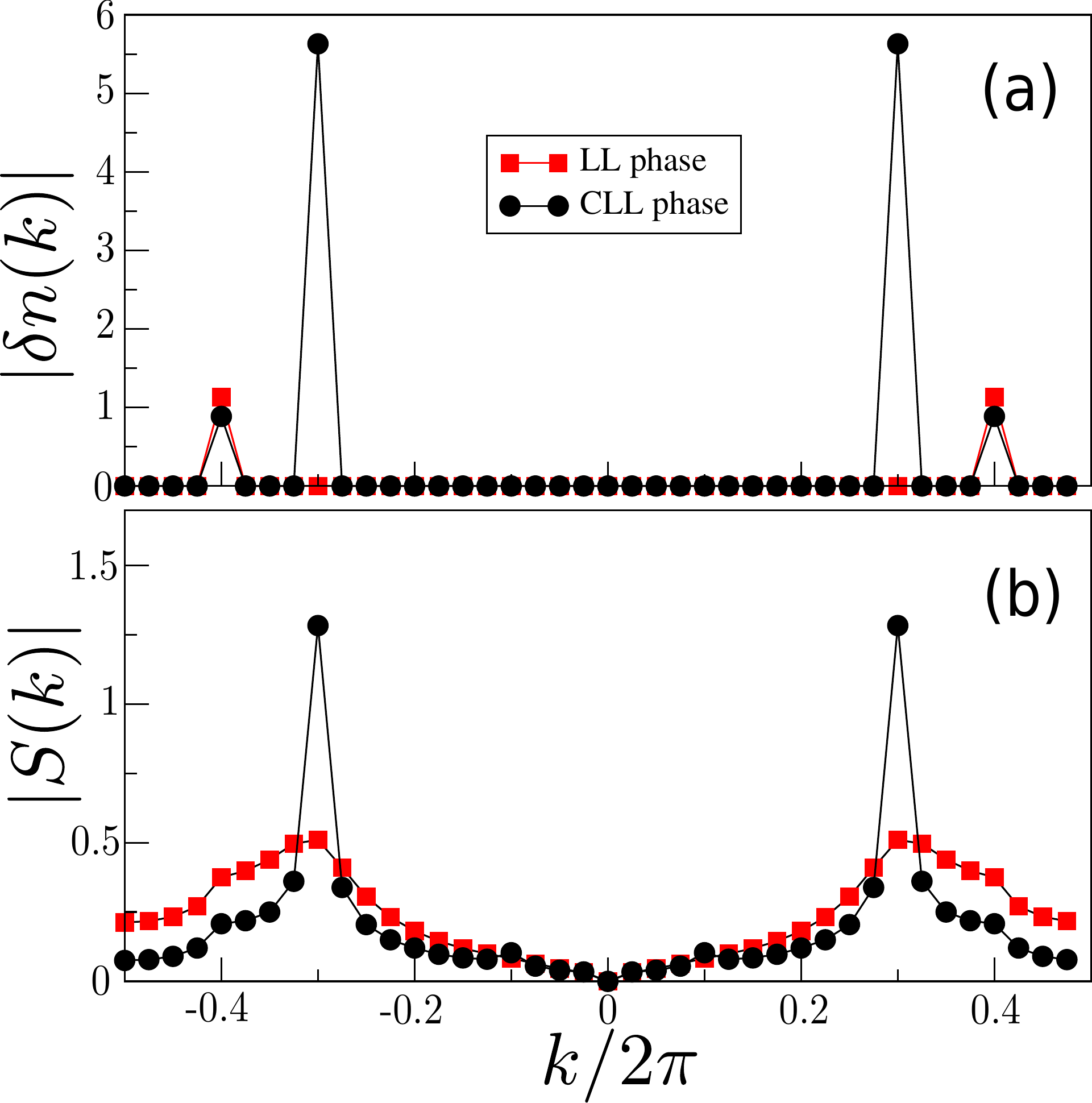}
\caption{DMRG results for $n=\frac{2}{5}$ along the $U=U_{2}=U_{1}$ line for $L=40$ and PBC. \textbf{\textsf{(a)}}
density fluctuations Fourier spectrum \eqref{density_fourier} and \textbf{\textsf{(b)}} density structure factor \eqref{structure_factor} both in the LL ($U=1$) and the CLL ($U=7$) phases. The momentum peak shift to the value $k=2\pi\cdot \frac{3}{10}$ is incompatible with a standard LL and supports that the physics of the CLL phase in the repulsive regime is ruled by the composite cluster degrees of freedom stemming from the classical limit analysis.}
\label{fig:nk_Sk_pos_04}
\end{figure}

In order to highlight the unconventional nature of the strong-coupling $c=1$ phase, we naturally investigate once more the spectral structure of the density profile and of the density-density correlation functions. 
We start by noticing that the classical limit ground state cluster density is given by $(1-n)/2 = \frac{3}{10}$, as one infers from the types of fundamental blocks relevant for the classical limit.
This offers a way to discriminate between the attractive CLL phase and the repulsive one by looking at the density dependence of the leading wavelength of the density fluctuations.

Our expectations are validated by the density profile spectrum exhibited in Fig.~\ref{fig:nk_Sk_pos_04}(a,b), where it appears manifestly that the density fluctuations are governed by the wave-vector $k=2\pi\cdot \frac{3}{10}$ above the transition point, consistently with the aforesaid classical limit argument and as opposed to the standard LL theory predictions. 
Analogous peaks appear in the static structure factor shown in Fig.~\ref{fig:nk_Sk_pos_04}, certifying thereby the irreducibility of the phase under investigation to the LL phase and the CLL phase of the attractive region.

\subsubsection{LL behaviour for $U_{2} < U_{1} / {2}$}

\begin{figure}[t]
\centering
\includegraphics[width=\columnwidth]{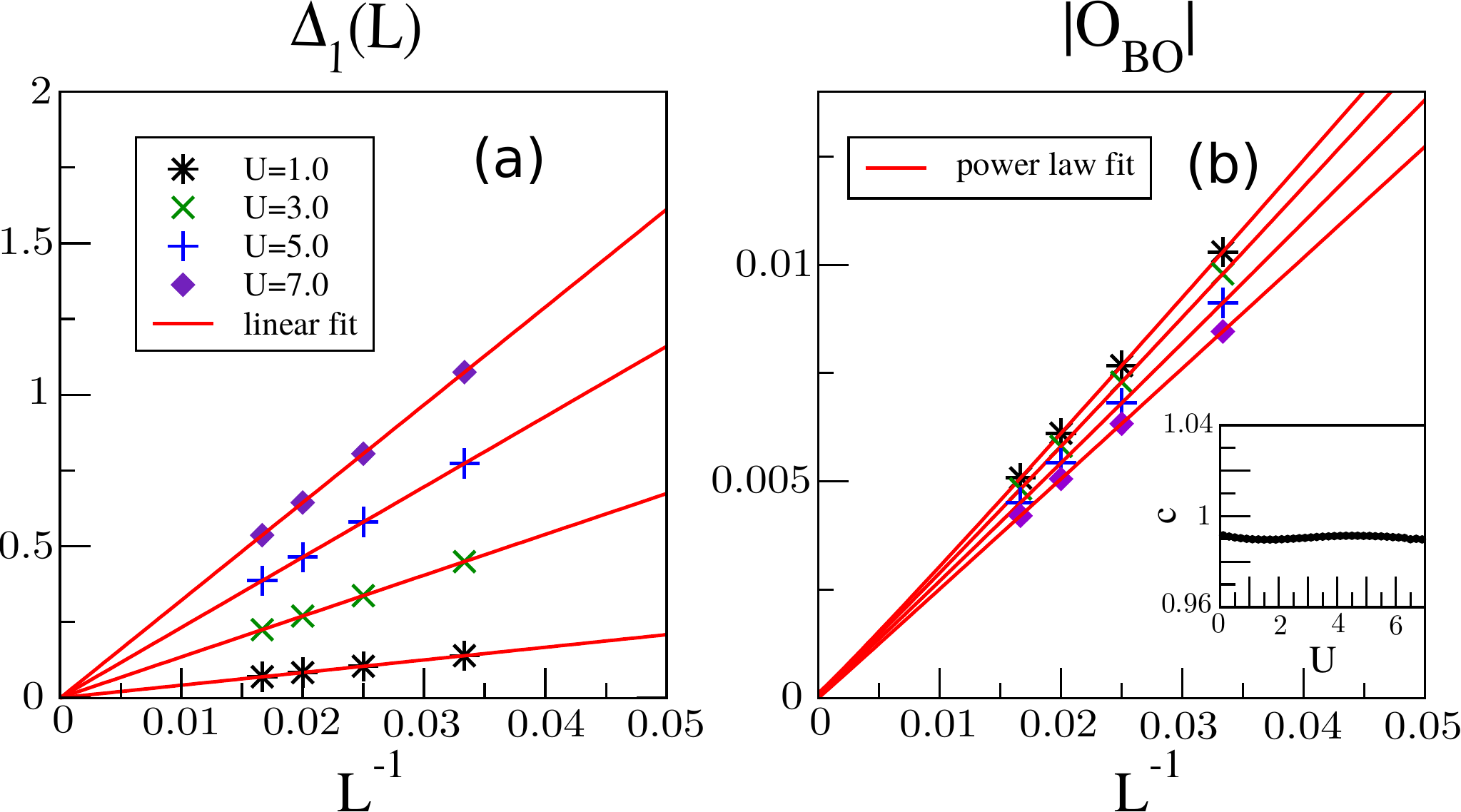}
\caption{DMRG results for $n=\frac{2}{5}$ along the $U=U_{2}=U_{1}/4$ line. \textbf{\textsf{(a)}}
Finite size scaling of the single-particle gap $\Delta_{1}(N,L)$ computed according to \eqref{gap}. A linear fit points towards zero extrapolated values  (three orders of magnitude smaller than the corresponding finite size values).
Such evidence of the gapless nature of single-particle excitations suggests the survival of the LL phase for a wide range of interactions.
\textbf{\textsf{(b)}} Finite size scaling of the BO parameter \eqref{bond_order}. Power law fits of the form $a_{0}+a_{1}L^{-a_{2}}$ are used and results in extrapolated values at least one order of magnitude smaller than the finite size values. Such behaviour further supports the LL picture.\textit{Inset:} Extrapolated central charge. The values deviate roughly at most $1.0\%$ from the $c=1$ value predicted for the LL phase.}
\label{fig:1gap_BO_c_pos_LL_04}
\end{figure}

We now consider the repulsive region defined by the condition $U_{2} < U_{1}/{2}$. We aim at verifying the conjecture of the survival of the weak-coupling LL phase in the whole region under consideration from numerical data obtained along the line $U_{2} = U_{1} /{4}$. 
We first consider the central charge, for which the results are collected in the inset of Fig.~\ref{fig:1gap_BO_c_pos_LL_04}(b) and do not hint towards a transition from the LL phase to any different phase. Indeed, the extrapolated central charge  deviates by at most $1\%$ from $c=1$.

Another conclusive evidence of the LL nature of this phase emerges from its spectral properties, as the single-particle gap displayed in Fig.~\ref{fig:1gap_BO_c_pos_LL_04}(a) scales to zero as the inverse system size for a wide range of values from weak- to strong-coupling, consistently with an adiabatic extension of the weak-coupling LL phase towards the strong-coupling regime.
Finally, a finite-size scaling analysis of the bound order parameter gives  the results displayed in Fig.~\ref{fig:1gap_BO_c_pos_LL_04}(b), where the extrapolated values go to zero on the scale of the finite-size sampled values, as expected to occur in a genuine LL phase.

\begin{figure}[t]
\centering
\includegraphics[width=0.85\columnwidth]{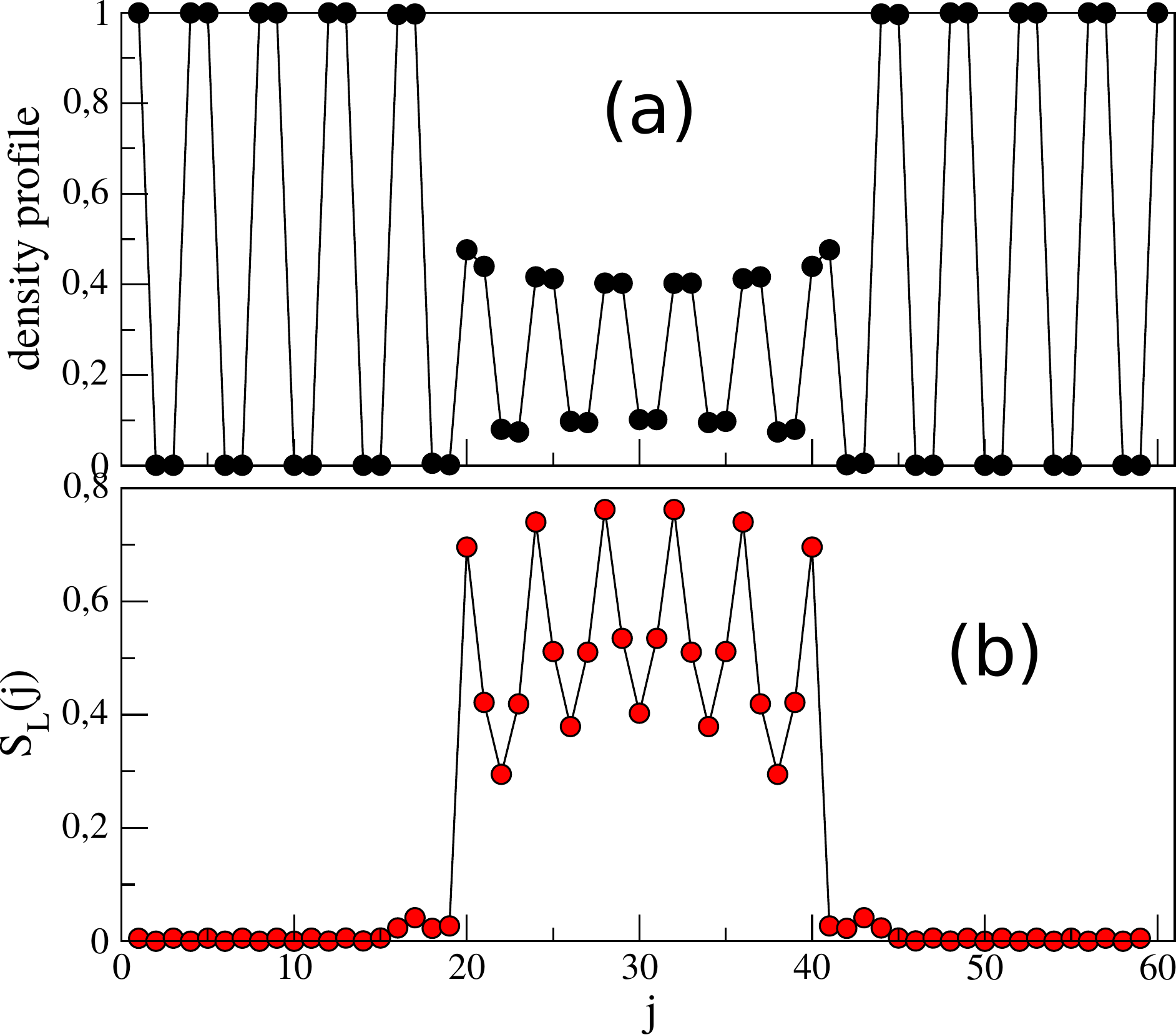}
\caption{DMRG results for $n=\frac{2}{5}$ along the $U_{1}=0$ line for $U_{2}=20$, $L=60$ and PBC. \textsf{\textbf{(a)}} Density profile. 
\textsf{\textbf{(b)}} Entanglement entropy profile $S_{L}(j)$.}
\label{fig:yellow_region}
\end{figure} 

\subsection{Phase separation along the $U_{1}=0$ axis} \label{yellow_region_subsection}

In order to give numerical evidence for the theoretical strong-coupling prediction along the $U_{1}=0$ axis, we provide the ground state expectation values of several observables at a point in parameter space where we expect the emergence of the corresponding phase. First, the density profile of Fig.~\ref{fig:yellow_region}(a) shows signatures of a phase separation with a region of perfect CDW$_{2}$ order and a confined liquid-like phase with strong residual oscillations in the density pattern typical of this large $U_{2}$ limit. 

Additionally, in order to confirm the strong coupling analysis, we remark that both the number of pairs in the CDW$_{2}$ region agrees with the  prediction for $N_{A}$ at density $n=\frac{2}{5}$ and the average density in the LL domain of the system, estimated as $\frac{1}{2}\left(\langle \hat{n}_{\frac{L}{2}-1}\rangle +\langle\hat{n}_{\frac{L}{2}}\rangle\right)$, coincides with its analytical estimate $n_{eff}=\frac{1}{4}$ apart from corrections of order $10^{-3}$ due to local quantum fluctuations.

As last indicator is the entanglement entropy profile is depicted in Fig.~\ref{fig:yellow_region}(b), where the CDW$_{2}$ domains feature negligible entropy, while the low density liquid-like region displays a finite entanglement.

\section{Phase diagram for $\mathbf{n=\frac 12}$} \label{fifth}

\begin{figure}[t]
\centering
\includegraphics[width=0.9\columnwidth,clip]{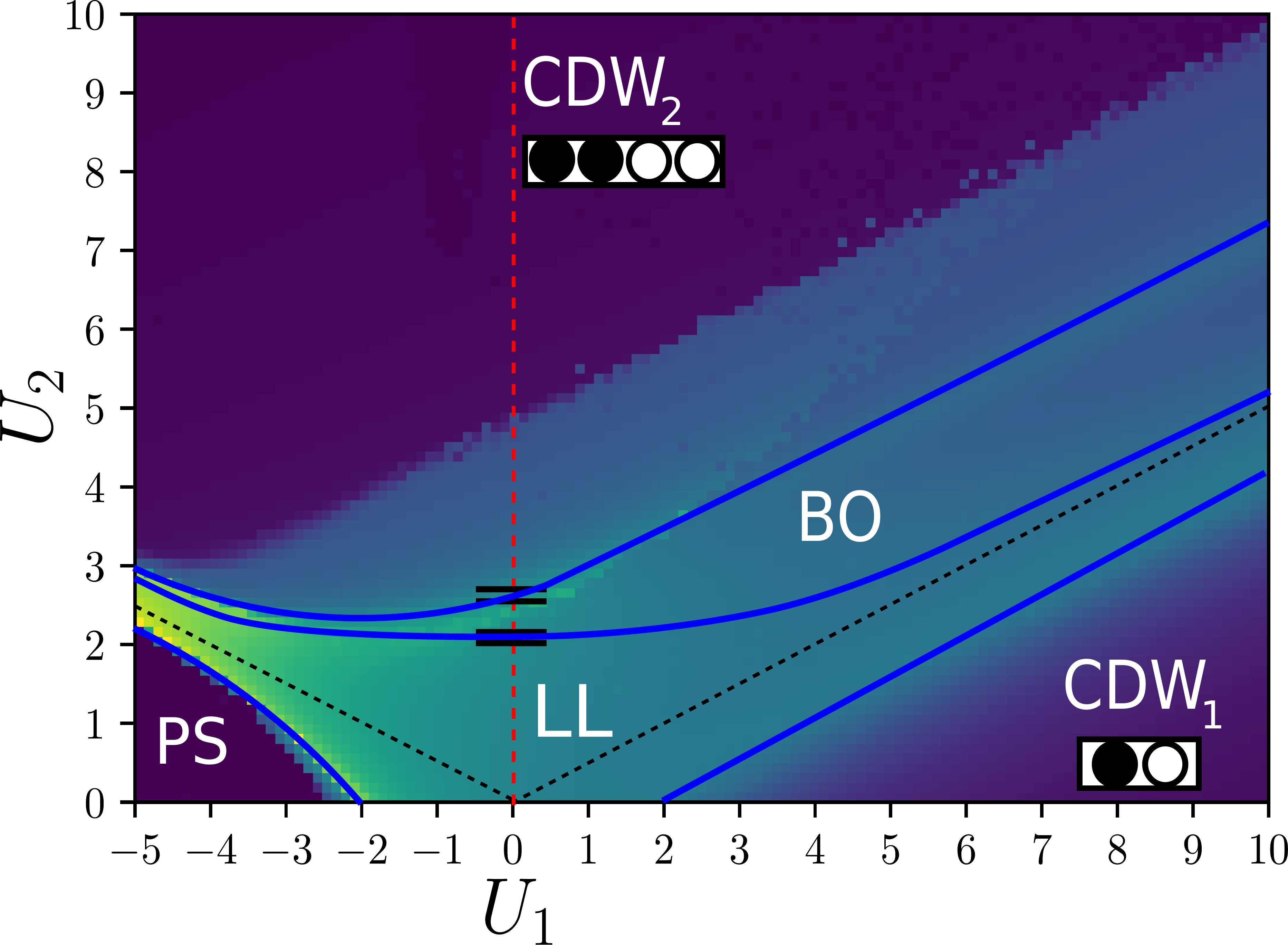}
\caption{Phase diagram for $n=\frac{1}{2}$. Color map for the background display the entanglement entropy on a $L=28$ chain with PBC. Black lines are classical transition lines obtained neglecting quantum fluctuation. Additional numerical simulations are presented for the points lying on the red lines. Blue lines are a guide to the eye for the main phase boundaries.}
\label{fig:PDn05}
\end{figure}

The following Section is devoted to the treatment of the $n=\frac 12$ case, for which the interplay between NN and NNN interaction terms and commensurability effects favor alternative orders such as two CDW orders and a BO phase (see Ref.~[\onlinecite{mishra_phase_2011}] for a thorough numerical characterization of the repulsive regime of model $\eqref{fermionic}$ at half-filling).
The global picture of the phase diagram of Fig.~\ref{fig:PDn05} is finally completed by characterizing the signatures of the transition to phase separation in the attractive NN interaction regime and the possibility to stabilize the CLL phase. 

\subsection{ Classical limit}

At half-filling, such a high density value favors formation of CDW states with no residual degeneracy, contrary to the configurations spaces at lower fillings. 
In order to rigorously derive such a result, we notice first of all that, while the phase separation limit remains, in strong coupling, in the region $U_{2}<-{U_{1}}/{2}$, the classical limit configurations giving rise to the CLL phases and comprised in between the lines $U_{2}=-{U_{1}}/{2}$ and $U_{2}= {U_{1}}/{2}$ simply turn into the periodic CDW$_2$ arrangement with unit cell ($\bullet\bullet\circ\circ$).
Indeed, the filling condition implies $N_{A}={L}/{4}$ and $N_{B}=0$, resp. $N_{C}=0$. 
In the same way, when $U_{2}<{U_{1}}/{2}$, the conditions $N_{D}={L}/{2}$ and $N_{C}=0$ indicate the appearance of the repulsive-NN-interaction-induced CDW configuration with unit cell ($\bullet\circ$). This suggests therefore the survival of such a CDW$_1$ phase, known to arise on the XXZ line, as far as the condition ${U_{2}}/{U_{1}} < {1}/{2}$ is fulfilled.

\subsection{Bosonization treatment}

We complete the analysis of the strong-coupling limit of the $n=\frac{1}{2}$ phase diagram with a weak coupling bosonization treatment of the Hamiltonian \eqref{fermionic}. 
The approach is indeed relevant mostly for $n=\frac{1}{2}$, where commensurability effects allow for the emergence of an Umklapp-induced term which in bosonization language reads as $H_{g}=g\int dx\cos\left(4\phi(x)\right)$, with the effective coupling $g\propto U_{2}-U_{1}$. The effective field theory of the low energy sector then takes the form of a sine-Gordon theory. At first order, the renormalization group (RG) flow equation that governs the evolution of the coupling $g$ reads as~\cite{giamarchi_quantum_2010}:
\begin{equation}
\frac{dg}{dl}=(2-4K)g\;,
\end{equation} 
which implies that, when $K<{1}/{2}$, the interaction term drives the system towards a non-liquid behaviour.

The latter instability is classified by introducing two order parameters, namely the local density fluctuations 
\begin{equation}
\delta n_{j}=\langle\hat{n}_{j}-\frac{1}{2}\rangle\sim\langle \frac{1}{\pi}\partial_{x}\phi+\frac{(-1)^{\frac{x}{a}}}{\pi a}\cos\left[2\phi (x)\right] \rangle\;,
\end{equation}
pointing towards CDW formation, and the local BO parameter 
\begin{equation}
B_{j}=(-1)^{j}\langle \hat{c}^{\dag}_{j}\hat{c}_{j+1}+h.c.\rangle\sim\langle\cos\left[2\phi(x)-\frac{\pi}{2}\right]\rangle\;,
\end{equation}
associated with the emergence of dimerization.
While the local density remains uniform, the local kinetic energy breaks translational invariance with alternating strong and weak bonds.
When $H_{g}$ is a relevant perturbation, the qualitative features of the resulting collective behaviour of the system is inferred by considering the limit $|g|\to\infty$, where the cosine term strongly locks the field $\phi(x)$ into the value that minimizes the interaction term.

If $U_{1}>U_{2}$, then $g<0$ and the field $\phi$ gets pinned at $\phi_{n}=n{\pi}/{2}$ with some integer $n$. As a result, $B_{j}\approx 0$ and $\delta n_{j}\propto (-1)^{j}$, thus capturing the emergence of the gapped CDW$_1$ phase with unit cell ($\bullet\circ$) that corresponds to the $U_2=0$ antiferromagnetic phase of the XXZ model.

If $U_{1}<U_{2}$, one has $g>0$ and the field $\phi$ gets pinned around the value $\phi_{n}=\frac{\pi}{4}+\frac{\pi}{2}n$ for some integer $n$. Consequently, $B_{j}\neq 0$ and $\delta n_{j}\approx 0$, hence providing evidence for a gapped dimerized phase with uniform density profile called BO phase, which is interpreted as the weak coupling precursor of the CDW$_2$ phase that has the unit cell ($\bullet\bullet\circ\circ$) and whose emergence 
has been predicted at strong coupling in the limit of dominant NNN interactions.

\begin{figure}[t]
\centering
\includegraphics[width=\columnwidth]{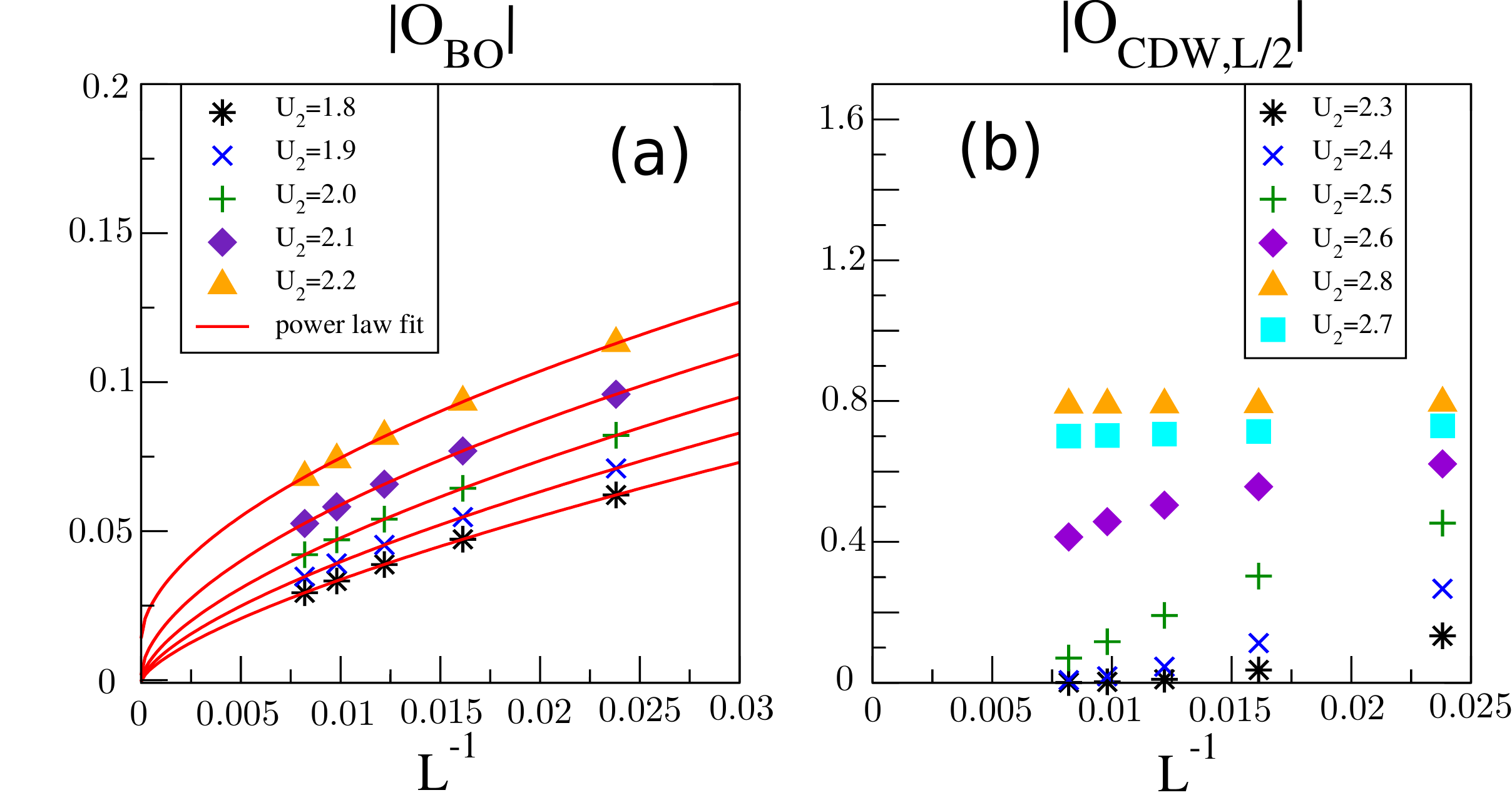}
\caption{DMRG results for $n=\frac{1}{2}$ along the $U_{1}=0$ line. \textbf{\textsf{(a)}}
Finite size scaling of the BO parameter \eqref{bond_order}. A power law fit of the form $a_{0}+a_{1}L^{-a_{2}}$ is used. 
The results show the onset of the BO regime, given that for $U_{2}=2.2$ the BO parameter extrapolates to a finite nonzero value of an order of magnitude larger than the extrapolated values for smaller $U_2$. \textbf{\textsf{(b)}} finite size scaling of the CDW order parameter \eqref{CDW_order}. The result demonstrates that the CDW order parameter scales to zero for $U_{2}<2.6$ but acquires a finite value for $U_{2}>2.6$. Thus, there exists an intervening BO phase.}
\label{fig:BO_CDW_05}
\end{figure}

\begin{figure*}[t]
\centering
\includegraphics[width=\textwidth]{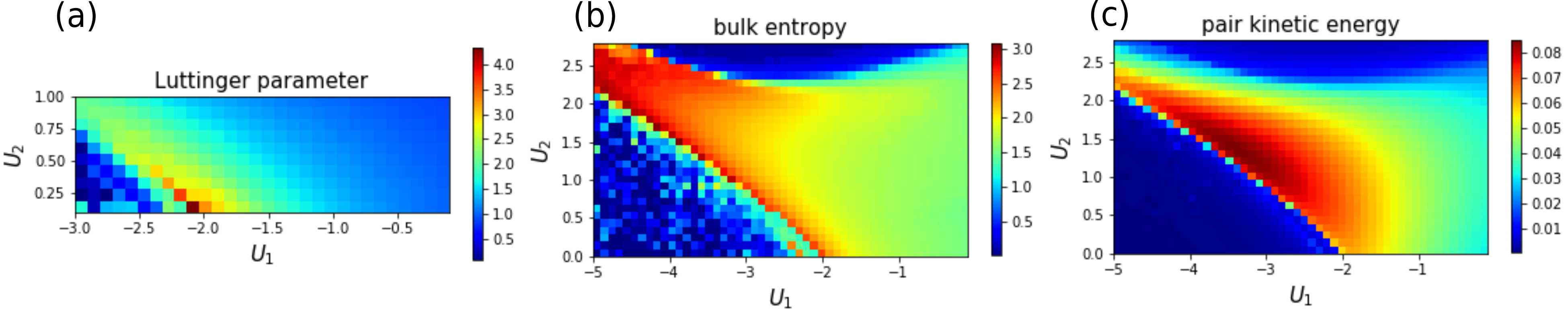}
\caption{DMRG results for $n=\frac{1}{2}$ for attractive $U_1$. 
\textsf{\textbf{(a)}}
Luttinger parameter $K$ obtained from fitting the pair correlation functions to a power law. It displays large values close to the PS phase, as it is known on the XXZ line showing that pair correlations are favored. [values inside the phase-separated region are meaningless and shown to visually identify the transition line].
\textsf{\textbf{(b)}} Entanglement entropy \eqref{entanglement_entropy} of half the system displaying the emergence of a characteristic peak separating the LL phase from phase separation and the appearance of the classical line $U_{2}=-{U_{1}}/{2}$ as the threshold above which phase separation turns into the ($\bullet\bullet\circ\circ$) CDW$_2$ configuration in the infinite coupling limit.
\textsf{\textbf{(c)}} Pair kinetic energy providing evidence for the enhancement of pairing fluctuations close to the transition line between the LL phase and phase separation, as expected from the divergent behaviour of the Luttinger parameter in the corresponding phase diagram region.}
\label{fig:K_05_neg}
\end{figure*}

\subsection{BO and CDW-II}

In order to benchmark the reliability of the weak-coupling bosonization predictions, we follow the BO parameter and the CDW-II order parameter along the $U_{1}=0$ axis. Our naive expectation consists in the observation of a first transition to a BO phase, in which the charge is localized on the bond connecting two neighbouring sites, followed by a successive transition to a CDW$_2$ with unit cell ($\bullet\bullet\circ\circ$), as predicted by  the classical limit analysis carried out at $n=\frac{1}{2}$.

The transition to the BO phase is probed by performing a finite size scaling analysis on the BO parameter as shown in Fig~\ref{fig:BO_CDW_05}(a). The result confirms the sudden appearance of a nonzero value of the BO order parameter, supporting the analytical predictions.
Furthermore, the BO phase is discriminated from the CDW$_2$ phase by means of the CDW$_2$ order parameter, which acquires a finite value in the thermodynamic limit for $U_2$ greater than the BO phase critical point, as demonstrated in  Fig~\ref{fig:BO_CDW_05}(b). It is worth noticing that, even though in the classical limit of the CDW$_2$ phase the BO parameter is expected to vanish, the latter still survives close to the BO-to-CDW phase transition due to residual kinetic fluctuations.

In the end, we are naturally lead to argue that, along all directions in between the lines $U_{2}=-{U_{1}}/{2}$ and $U_{2}={U_{1}}/{2}$ (having the  CDW$_2$ configuration as the classical limit), an intermediate, stripe-shaped BO phase region intervenes between the LL phase and the CDW$_2$ phase.
It has a width which decreases as the abrupt classical limit transition between the CDW$_2$ and the other classical limit configurations (CDW$_1$ and phase separation) is approached.

\subsection{Phase separation}

Last, we focus for $n=\frac{1}{2}$ on the vicinity of the phase separation transition line because we expect it to favor pairing fluctuations in the LL regime.
First, as we have seen, phase separation occurs whatever the density for $U_{1}=-2.0$ on the $U_{2}=0$ XXZ line.
By adding a positive $U_{2}$, we expect that the transition line is shifted  to a more negative $U_{1}$ as the repulsive NNN interaction term increases. Indeed, its effect will consist in the destabilization of the phase-separated macroscopic cluster by means of an additional $O(N)$ contribution to its mean energy.

A first indication of the correctness of our interpretation of the behaviour of the critical line separating the LL phase from the phase-separated one is obtained by looking at the behavior of the Luttinger parameter $K$.
As known from the Bethe Ansatz solution of the XXZ model, $K$ diverges close to the isotropic Heisenberg point according to the formula~\cite{luther_calculation_1975}:
\begin{equation}
K=\frac{\pi}{2(\pi-\arccos\Delta)},
\end{equation}
where $\Delta$ is the anisotropy parameter of the XXZ chain. Hence, we decide to fit the Luttinger parameter from the pair correlator decay in the region of interest, expecting its divergent behaviour to mark the approximate location of the transition. The result is presented in Fig~\ref{fig:K_05_neg}(a), where the largest $K$ values are achieved approximately in correspondence of the entropy peak.

Indeed, monitoring the bulk entropy magnitude as shown in Fig.~\ref{fig:K_05_neg}(b) nicely shows the limits of the LL regime between the low-entropy gapped phases and phase-separated phase.
It is clearly bending towards smaller $U_{1}$ values as $U_{2}$ increases, is signaled by a bump in the bulk entropy value, consistently with the behaviour of the entanglement entropy in Heisenberg-like models predicted in Ref.[\onlinecite{popkov_logarithmic_2005}].

Since a large value of the Luttinger parameter is associated to an enhancement of pairing fluctuations, we finally present in Fig.~\ref{fig:K_05_neg}(c) the pair kinetic energy, which gets significantly larger when approaching the transition to phase separation, consistently with the previous findings.

\section{Conclusions}\label{Sec:Conclusion}

The present work proposes a comprehensive description of the ground-state phase diagram of Hamiltonian~\eqref{fermionic} as a function of the density, highlighting the emergence of exotic phases departing from the LL paradigm. By means of entropic, spectral, and correlation properties, their most notable signatures have been unveiled and benchmarked against the results of a classical-limit analysis and of effective field-theory treatments.

The topology of the phase diagram in the attractive $U_{1}<0$ region is observed to exhibit a strong robustness against the variation of density for $n< \frac 12$. It features phase separation and, more importantly, a CLL of pairs, separated from the LL phase by a $c=\frac{3}{2}$ critical point and characterized in a phenomenological way (i) by the opening of a gap in the single-particle excitations and (ii) by anomalous peaks in the Fourier spectrum of various observables. 

% The only qualitative change occurs at $n=0.5$, where the CLL phase is replaced by a CDW of localized pairs.

The phenomenology in the repulsive $U_1>0$ regime has been shown to exhibit a much richer behaviour. Firstly, as shown by finite-size scaling analyses on the central charge and on the BO parameter $\eqref{bond_order}$, the LL phase appears to be the only zero-temperature phase of the model when the density satisfies $n<\frac{1}{3}$. 
On the other hand, for $n=\frac{1}{3}$, the interplay between the interaction range and the density induces a transition from the LL phase to a gapped, strong-coupling CDW phase whose classical limit configuration exhibits one particle every three sites. For this phase, we have presented the standard characteristic signatures of a gapped crystalline phase.
In the density range $n \in (\frac{1}{3},\frac{1}{2})$, the fine-tuning of the density which gives rise to the CDW phase is removed and one observes a transition to a frustration-induced CLL phase. We have discriminated the latter from its attractive regime counterpart by means of the peak location in the density-fluctuation Fourier series and in the structure factor, which in turn we have related to the structure of the corresponding classical-limit ground state. On top of the above considerations, we have predicted and characterized the phase-separated regime emerging at large $U_{2}$ and small $U_{1}$ in the density range $[\frac{1}{3},\frac{1}{2})$, thereby theoretically justifying the numerically probed coexistence of liquid and CDW$_2$ orders in the two macroscopic phase domains of the system. 

As expected, when the model is studied at half-filling,
both attractive and repulsive sides of the phase diagram are significantly modified. In particular, at strong coupling, the aforementioned liquid phases are replaced by CDW phases whose structure is dictated by the dominant interaction term inducing them. Additionally, we predicted by means of bosonization calculations and confirmed numerically the appearance of a BO phase at intermediate coupling, featuring localization of the fermionic particles on the bond connecting neighbouring sites rather than on a single site, as in the case of the strong-coupling CDW counterparts.

Put in a broader perspective, we expect this work to guide the search for Majorana fermions in inhomogenous fermionic systems composed of paired and normal fluids. Additionally, it could help revealing paired phases in upcoming experiments with Rydberg atoms, where long-range interactions cannot be neglected. The large tunability in terms of excitation densities and interaction strength of those setups makes it reasonable that several regimes of the model studied in this article could be experimentally accessed.

\acknowledgments

We acknowledge funding by the Agence Nationale de la Recherche (ANR) under the project TRYAQS (ANR-16-CE30-0026) and by LabEx PALM (ANR-10-LABX-0039-PALM). This work has been supported by Region Ile-de-France in the framework of the DIM Sirteq.

%\nocite{*}

%\bibliographystyle{apsrev4-2}
%\selectlanguage{english}
\bibliography{paper.bib}

% one must clean up the bib file, some names make it bug

\end{document}